\documentclass[structabstract]{aa}
\usepackage[modulo,switch]{lineno}
\usepackage{graphicx}
\usepackage{natbib}
\usepackage{longtable}
\usepackage{longtable,lscape}
\usepackage{txfonts}
\begin{document}
\linenumbers
\newcommand{\meandnu} {\langle\Delta\nu\rangle}
   \title{Solar-like oscillations in red giants observed with \textit{Kepler}: influence of increased timespan on global oscillation parameters\thanks{Values of the global oscillation parameters can be obtained from the authors upon request.}}

   \author{S. Hekker\inst{1,2} \and Y. Elsworth\inst{2}  \and B. Mosser\inst{3} \and T.Kallinger\inst{4} \and W.J Chaplin\inst{2} \and J. De Ridder\inst{4} \and R.A. Garc\'ia\inst{5} \and D. Stello\inst{6} \and B.D. Clarke\inst{7} \and J.R. Hall\inst{8} \and K.A. Ibrahim\inst{8}}

\offprints{S. Hekker, \\
                    email: S.Hekker@uva.nl}
   \institute{Astronomical institute `Anton Pannekoek', University of Amsterdam, Science Park 904, 1098 XH, Amsterdam, the Netherlands
   \and University of Birmingham, School of Physics and Astronomy, Edgbaston, Birmingham B15 2TT, United Kingdom
   \and LESIA, UMR8109, Universit\'e Pierre et Marie Curie, Universit\'e Denis Diderot, Observatoire de Paris, 92195 Meudon Cedex, France
    \and  Instituut voor Sterrenkunde, KU Leuven, Celestijnenlaan 200D, 3001 Leuven, Belgium
     \and Laboratoire AIM, CEA/DSM-CNRS, Universit\'{e} Paris 7 Diderot, IRFU/SAp, Centre de Saclay, 91191, GIf-sur-Yvette, France
     \and Sydney Institute for Astronomy (SIfA), School of Physics, University of Sydney, NSW 2006, Australia
     \and SETI Institute/NASA Ames Research Center, Moffet Field,, CA 94035, USA
     \and Orbital Sciences Corporation/NASA Ames Research Center, Moffet Field, CA 94035, USA \\
         }

   \date{Received ; accepted}

\abstract
   {The length of the asteroseismic timeseries obtained from the \textit{Kepler} satellite analysed here span 19 months. \textit{Kepler} provides  the longest continuous timeseries currently available, which calls for a study of the influence of the increased timespan on the accuracy and precision of the obtained results.}
   {We aim to investigate how the increased timespan influences the detectability of the oscillation modes, and the absolute values and uncertainties of the global oscillation parameters, i.e., frequency of maximum oscillation power, $\nu_{\rm max}$, and large frequency separation between modes of the same degree and consecutive orders, $\meandnu$.}
   {We use published methods to derive $\nu_{\rm max}$ and $\meandnu$ for timeseries ranging from 50 to 600 days and compare these results as a function of  method, timespan and  $\meandnu$.}
   {We find that in general a minimum of the order of 400 day long timeseries are necessary to obtain reliable results for the global oscillation parameters in more than 95\% of the stars, but this does depend on $\meandnu$. In a statistical sense the quoted uncertainties seem to provide a reasonable indication of the precision of the obtained results in short (50-day) runs, they do however seem to be overestimated for results of longer runs. Furthermore, the different definitions of the global parameters used in the different methods have non-negligible effects on the obtained values. Additionally, we show that there is a correlation between $\nu_{\rm max}$ and the flux variance.}
   {We conclude that longer timeseries improve the likelihood to detect oscillations with automated codes (from $\sim$60\% in 50 day runs to $>$ 95\% in 400 day runs with a slight method dependence) and the precision of the obtained global oscillation parameters. The trends suggest that the improvement will continue for even longer timeseries than the 600 days considered here, with a reduction in the median absolute deviation of more than a factor of 10 for an increase in timespan from 50 to 2000 days (the currently foreseen length of the mission). This work shows that global parameters determined with high precision - thus from long datasets - using different definitions can be used to identify the evolutionary state of the stars.}

   \titlerunning{Influence of increased timespan on global oscillation parameters}
   \authorrunning{Hekker et al.}
   \maketitle
%

\section{Introduction}
Many breakthrough results for red-giant (G-K) stars have been presented using data obtained by the CoRoT \citep{baglin2006} and NASA \textit{Kepler}  \citep{borucki2010} missions. These results include statistical ensemble studies of global oscillation parameters, i.e., frequency of maximum oscillation power, $\nu_{\rm max}$, mean frequency separation between modes of the same degree and consecutive orders, $\meandnu$, small frequency separations between modes of different degree, $\ell$, amplitudes and visibilities of the oscillations, and tests of scaling relations \citep[e.g.,][]{deridder2009,hekker2009,bedding2010,huber2010,hekker2011pub,huber2011,mosser2012}. Additionally, it has been possible to determine stellar parameters such as masses and radii \citep{kallinger2010corot,kallinger2010}. In addition to these results, asteroseismic investigations into the granulation \citep{mathur2011}, red giants in clusters \citep{basu2011,hekker2011clus,stello2011mem,stello2011ampl} and red giants in eclipsing binaries \citep{hekker2010ecb} have been performed, as well as detailed investigations into the internal structure of single stars \citep[e.g.,][]{dimauro2011,jiang2011,baudin2012}.
The \textit{Kepler} results referred to are based on timeseries with a near regular cadence of either 29.4 min or 58.85 s  and a timespan ranging from $\sim$30 days up to more than 1.5 yr. These are the first datasets from space-based telescopes with such long timespan and high fill ($\gtrsim$90\%) and frequency resolution ($\approx$ 0.019~$\mu$Hz). Underpinning much of this work is the ability to determine global oscillation parameters and the uncertainties in these values. It is reasonable to ask if there are now enough data available and whether there are any gains to be obtained from observing individual stars for longer periods.
In this paper we address the precision and reliability of the determination of some of the global seismic parameters.
There are other areas where there is a clear need for data of longer duration because the features detected in the power spectra are narrow and hence barely resolved even by the current datasets. In particular, we highlight the detection of  g-p mixed modes \citep{beck2011}. The observed mean period spacings appear to have different values for stars that  burn only H (in a shell) and those that also burn He in the core \citep{bedding2011,mosser2011mm}, hence the period spacing can be used to distinguish between different evolutionary states in which red giants are observed using the characteristics of their frequency spectra. Another method to distinguish between different evolutionary phases is based on the difference in frequency dependence of radial modes \citep{kallinger2012}. Furthermore, recently, the timeseries obtained with \textit{Kepler} have become long enough to study rotational splitting of the oscillation modes, which led to the detection of differential rotation in red giants \citep{beck2012}.

In this work, we use the 19 months of data available from Q0 to Q7 to investigate how the increased timespan influences the detectability of the oscillation modes, and the absolute values and uncertainties of the global oscillation parameters, $\nu_{\rm max}$ and $\meandnu$. These are important in several ways. Knowing the dependence of the precision on data duration is a guide for observing strategies, and for the determination of those secondary parameters that are derived from the primary global oscillation parameters, such as stellar mass and radius.  Furthermore, it is crucial to be able to estimate the proportion of false negatives and false positives for population studies. Also, for detailed modelling of individual oscillation frequencies $\nu_{\rm max}$ turned out to be of great diagnostic potential \citep{gruberbauer2012}. We will include in our considerations the impact of other relevant parameters such as the observed height-to-background ratio of the oscillation excess. This work is a follow-up of  \citet[][hereafter paper I]{hekker2011comp} on the red giants and \citet{verner2011} on solar-type stars, in which results obtained with different methods have been compared and validated.

Paper I described the comparison of global oscillation parameters extracted from about four month of \textit{Kepler} data using different methods. From this comparison, it was concluded that
1) the results from the different methods agree for most stars within a few percent;
2) at least five methods (out of the seven tested)  obtained results for 92\% of stars for $\nu_{\rm max}$ within the range of 50~$\mu$Hz to 170~$\mu$Hz, and this percentage decreased to 69\% when  all stars with $\nu_{\rm max}$ covering the complete frequency range, i.e., 0 -- 283.4\,$\mu$Hz (the Nyquist frequency) were included;
3) the scatter due to realization noise, originating from the stochastic nature of the oscillations, is non-negligible and can be at least as important as the internal uncertainty of the results due to the method used, but this depends on the frequency of maximum oscillation power, $\nu_{\rm max}$, and on the methods. In case a model is used to describe the variation of $\Delta\nu$ with frequency the results are less sensitive to realization noise than others; 
4) the influence of the obtained value of $\meandnu$ is less dependent on the frequency range over which it is computed than is the case for solar-type stars. A theoretical follow-up study to explain the latter has been performed by \citet{hekker2011dnu}.

\section{Data}
For the current study, we use data obtained with the \textit{Kepler} satellite during its first $\sim$19 months of operation (Q0-7). These data have a $\sim$29.4 minute near regular cadence and have been corrected for possible artifacts in the way described by \citet{garcia2011}. See e.g. \citet{jenkins2010} for some characteristics of these data. The stars in the sample investigated here have been selected for asteroseismic investigations by the \textit{Kepler} Asteroseismic Science Consortium (KASC) or for astrometric purposes. We exclude cluster stars  from this sample. Additionally, we include only stars for which a power excess characteristic for stochastic oscillations is detected.
In some cases the stars episodically fall on the one CCD that has gone inactive, resulting in loss of data. We exclude these stars from our current investigation.  Other causes of data loss are safe mode and momentum dumping from the spacecraft, as well as data downlinks every $\sim$30 days. These result in rather smaller losses of data.  We require that the stars have been observed in all available quarters and we accept a fill level down to 94\% accounting for some additional loss of data. This then leaves us with 1028 stars.

\begin{figure}
\begin{minipage}{\linewidth}
\centering
\includegraphics[width=\linewidth]{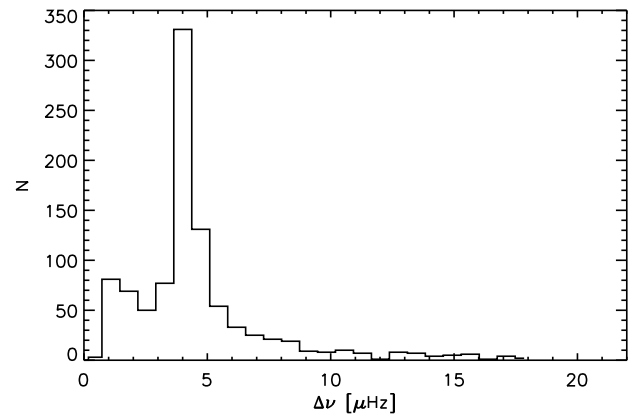}
\end{minipage}
\caption{Distribution of the mean large frequency separations of  the stars in our sample.}
\label{dnu-distribution}
\end{figure}

The $\meandnu$ distribution of stars in the dataset we consider here is shown in Fig.~\ref{dnu-distribution}. This distribution is similar to the ones seen in other published work on the \textit{Kepler} red giants \citep[e.g.,][]{hekker2011pub}.

\section{Parameter extraction}

For the data analysis, all the methods used here are based on a subset of those described in paper I,
i.e. COR \citep{mosser2009,mosser2011}, OCT \citep{hekker2010method} and CAN \citep{kallinger2010}.
For $\nu_{\rm max}$ values we have used the autocorrelation function from COR:EACF \citep{mosser2009}, and
the centre of the Gaussian fit to the oscillation power excess from OCT \citep[method II in][] {hekker2010method} and CAN \citep{kallinger2010}.
For $\meandnu$ we use the autocorrelation method \citep[COR:EACF,][] {mosser2009} and the universal pattern \citep[COR:UP,][]{mosser2011} as well as the determination of the peak in the power spectrum of the power spectrum using statistics of grouped data OCT:PS$\otimes$PS and with the addition of  bayesian statistics OCT:PS$\otimes$PS (bayesian) \citep{hekker2010method}, and finally, fitting of the central three radial orders \citep[CAN,][]{kallinger2010}.


A homogeneous comparison between the values of the shorter timeseries as presented in paper I, and of longer timeseries cannot be performed directly, as continuous improvements to the methods have been made. These improvements have been made as a result of our increasing knowledge of the data from earlier runs and to deal with the longer timeseries. The changes are of numerical nature and do not alter the underlying principles of the methods. Hence, the references cited above are still valid.
To perform a uniform study of the impact of the length of the timeseries, the (Q0-Q7) dataset ($\sim$600~days) has been used both  as a whole and divided into subsets. These datasets are all analysed with the latest versions of the analysis methods.

\section{Likelihood of detecting oscillation power in frequency spectra}
Recently, \citet{hekker2011pub} analysed one-month data sets of publicly available data for over 16\,000 red giants selected on the basis of effective temperature and surface gravity. They  found that in $\sim$70\% of the stars, oscillations could be detected. This raises questions as to whether this fraction is telling us something about the ability of red giants to sustain stochastically-driven oscillations, or if it is just a reflection of the difficulties in the automated detections of oscillations for relatively short data sets?  Perhaps also, some of the stars were so faint that their noise levels prevented the oscillations being detected. Alternatively, can some other feature in the star  suppress the oscillations in the same manner as activity is known to suppress the oscillations in solar-like stars \citep{mosser2009act,chaplin2011act,huber2011}? Here we will first give consideration to the importance of the amplitude of the oscillations and apparent brightness of the stars and we will subsequently consider the problems associated with the automated methods.

We consider how we might estimate the likelihood of detecting oscillation power when it is present in the data. We use the same method as given in  \citep{chaplin2011detect}, adapted for the red giants,  to show that there is a high expectation that we will be able to detect the modes of oscillations in all the red giants in the \textit{Kepler} data set.
It is important to note that, although the correct identification of the frequency range in which the modes are located is of fundamental importance, most current methods do not use this as their primary consideration when determining if there is oscillation power in the spectrum. For most of the methods, the determination of $\Delta \nu$ is done first. If this fails then `no detection' is reported. This may not be the best strategy, but before we construct that discussion we should first explore the existing predictions for the amplitudes of the modes of oscillations in red giants.

\subsection{Prediction for mode power}
 \citet{kjeldsen1995} devised scaling relations predicting that the amplitude of solar-like oscillations scale with their luminosity to mass ratio, which implies that the amplitude of the oscillations increases with increasing stellar radius. Hence, solar-like oscillations in red giants are expected to have higher amplitudes than oscillations in solar-type stars of equal masses. These scaling relations have recently been revised \citep{kjeldsen2011}, and also tested, both theoretically \citep[e.g.][]{samadi2007} and observationally \citep[e.g.][]{baudin2011,baudincor2011,huber2011,stello2011ampl}.

 To determine if it is possible to detect the modes, we are interested in the signal-to-noise ratio in the vicinity of the modes.
 In \citet{mosser2012} it was shown that, with a small dependence on evolutionary status, the ratio of the height of the smoothed power spectrum to the granulation noise background evaluated at $\nu_{\rm max}$ is between 3.7 and 4.0 for clump and red-giant branch stars, respectively. Accordingly, we will use the lower limit of this to work out the signal-to-noise in the integrated spectral power excess.
 For all the red giants that we consider here, the intrinsic photon shot noise is negligible and we neglect it. This removes a consideration of the stellar luminosity from the calculations.

 A commonly accepted model of the envelope of the oscillation power is a Gaussian function whose width, $W_{\rm env}$, scales with the frequency of maximum power $\nu_{\rm max}$ as
 $W_{\rm env} = 0.59\nu_{\rm max}^{0.9}$  \citep{mosser2010}.
We can determine the average power in the oscillations by smoothing the power spectrum over a range of at least one large spacing so that no trace of the individual modes remains. It is recognized that doing this in practice requires considerable care as is spelt out in \citet{mosser2012}.
We will take twice the full-width half-maximum of the underlying Gaussian as the range over which we will integrate to determine the average power. This range  contains all but a few per cent of the oscillation power.
The granulation background in the vicinity of the modes can be modeled with a power law with index of $-2.1$ \citep{mosser2012}. Integration of these two functions, over the same range, leads to an integrated height-to-background ratio ($H/B_{\rm int}$ ) of 1.55 (0.42 the height-to-background ratio at $\nu_{\rm max}$).

The averaging of the data during the sampling interval in the time domain causes an attenuation of the amplitudes in the frequency spectrum according to a sinc function and we can use the ratio of $\nu_{\rm max}$ to $\nu_{\rm Nyq}$, the Nyquist frequency which is 283.4~$\mu$Hz for the \textit{Kepler} long cadence data, to quantify the size of this reduction.
The majority of stars in our sample have $\nu_{\rm max} \ll \nu_{\rm Nyq}$ and for these stars this sinc term is negligible and we do not consider it further.

\subsection{Model of detection probability}
\label{ss:detection-prob}
Here we present a model, based on predicted integrated height-to-background in the vicinity  of the oscillations, for how detectable the oscillations are. To do this we adapt the formulation devised by \citet{chaplin2011detect} for solar-type stars to red giants.
The principle of the method is to compare the power present in the modes with that present in the background and then to use probability distributions to ascertain the likelihood of the mode power being reliably detected.

\begin{figure*}
\begin{minipage}{\linewidth}
\centering
\includegraphics[width=\linewidth]{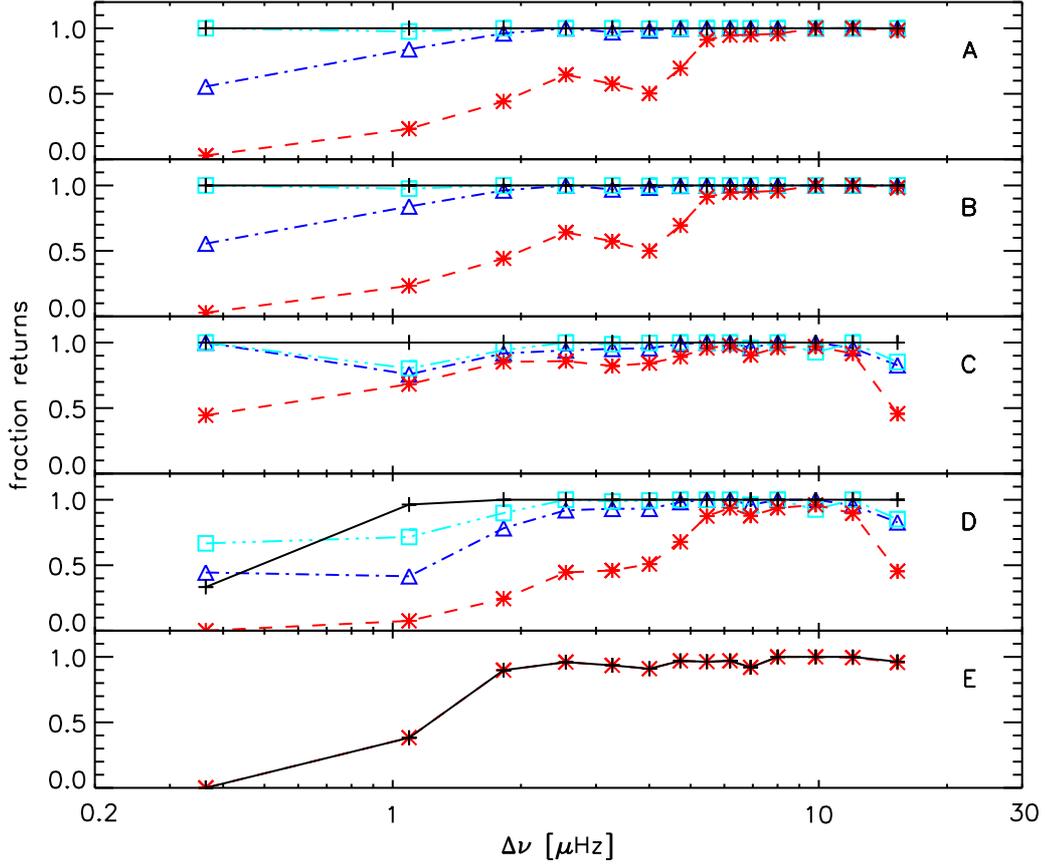}
\end{minipage}
\caption{Fraction of runs with returned values for each star per $\Delta \nu$ interval. Each panel shows the results of a certain method (A: COR - Universal Pattern, B: COR - EACF, C: OCT - PS$\otimes$PS, D: OCT - PS$\otimes$PS (bayesian), E: CAN) with run length 50, 200, 400, 600 days in red, blue, cyan and black, respectively. Note that the 50 and 600 day curves in panel E overlap due to the fact that the 600 day results were used to constrain the input for the 50 day runs. No results for 200 and 400 day long runs were obtained by CAN.}
\label{returnhisto}
\end{figure*}

The question that we now wish to answer is `given the $H/B_{\rm int}$ what is the chance of a false
detection?'. We set a probability, $p_{\rm false}$, at which we are prepared to risk a false positive detection.
In general, this level should be low. Typically for this work we have used $p_{\rm false}$=0.01 (i.e.1\%).
As detailed in  \citet{chaplin2011detect}, we compute a threshold value, $\theta$, in a $\chi ^2$ distribution with $2n$ degrees of freedom (dof) such that the probability that some random variable is greater than the threshold value supplied is equal to $p_{\rm false}$. In this, we take $n$, the number of degrees of freedom, as the number of independent frequency bins used to compute $H/B_{\rm int}$.  We must also take account of the chance that, because of random noise in the data, we will miss a true detection for a star with sufficient signal-to-noise for detection which leads to a new threshold value $\theta_2$.

\begin{equation}
\theta_2=\frac{\theta+1} {H/B_{\rm int}+1}.
\label{eq3}
\end{equation}
This value $\theta_2$ is then used to derive probability $p$, where $p$ is the probability that in a $\chi^2$ distribution with $2n$ dof a random variable is less than or equal to the cut off level specified.
Finally we have the probability we sought which is $p_{\rm final} = 1-p$
the probability that a given $H/B_{\rm int}$ will exceed the  computed threshold $\theta$.

The recipe as described predicts that for all stars considered here (even for datasets as short as 50~days) we are likely to detect the oscillations. In general, the lower probabilities are at about 93\% likelihood for stars which have $\nu_{\rm max}$ below 10~$\mu$Hz. For one particular star the detection probability dropped to 75\%. 
Note that these predictions are not sensitive to the shape of the oscillation power excess nor to any structure, such as the large frequency separation in it, and that we have taken the worst case scenario for the $H/B$ of Helium-core-burning evolutionary status. These results are based on the integrated power of the oscillations. So from this test it appears that a detection rate higher than 70\%  as quoted by \citet{hekker2011pub} would be expected when using the $H/B$ indications. But how does this compare with observational results from longer timeseries? We now consider this issue in the next section. 

\section{Observational results}

For observed stars we do not know the true values of the seismic parameters. All that we can do is to estimate them using the observations. In order to obtain such estimates of the seismic parameters $\nu_{\rm max}$ and  $\Delta\nu$, the COR and OCT methods are used to analyse the full timespan of just under 600~days of the complete set of stellar data. The analysis was `blind' in that no manual checks were made on the outcomes. We therefore expect to have some errors in the results. We did not use the CAN method because, for computational reasons (the multinest procedure is very time consuming), not all available stars were analysed with it.
For 974 stars there is close agreement between the results from OCT and COR for $\nu_{\rm max}$ and $\meandnu$. In this context, close agreement is taken to be that the two completely independent methods identify the oscillations in the same region of the spectrum to within half the expected width of the envelope of the oscillation power, with the width of the oscillation envelope as  defined by \citet{mosser2010}. Taking this relatively relaxed constraint is justified by the fact that we want to select a statistically significant sample of stars with oscillations detected by different methods in the same frequency range. For the remaining 54 stars, there are disagreements between the values obtained with the different methods. We inspected these stars by eye and for 39 stars the oscillations are at low frequencies ($\nu$~$<$~5~$\mu$Hz), for four stars the oscillations straddle the Nyquist frequency and for 11 stars we do not have the standard red-giant oscillation spectrum due to the presence of artefacts or these could possibly be mis-classified as red giants.

For the 974 stars for which there is agreement, we create reference values which are the mean values of $\nu_{\rm max}$ and $\meandnu$, respectively. These reference values are essentially an arbitrary zeropoint used to select reliable results and to discard outliers.

\subsection{Outlier removal in short datasets}
When short datasets are considered there will be occasions when the returned values are unreliable. We wish to remove some of these so that we can look at the spread in the reliable results. A very simple outlier rejection algorithm is used whose purpose is to reject patently wrong answers. This is the same as described in paper I and depends on comparing the reference value with the individual values. The results presented in \citet{verner2011} suggest that for solar-type stars it is appropriate to use rejection criteria that scale with the $\nu_{\rm max}$ of the star. However, it was shown in paper I that this is not appropriate for red-giant stars. The process adopted here first rejects points that are more than 50\% different from the reference value, irrespective of $\nu_{\rm max}$ or $\meandnu$, and then applies an absolute cut. For these absolute cuts a value of 10~$\mu$Hz has been used for all but the low values of  $\nu_{\rm max}$ and a cut of 2~$\mu$Hz has been used for $\meandnu$. The cross-over position where the absolute cut off is more stringent than the relative one occurs at about $\nu_{\rm max}$~=~20~$\mu$Hz.

\subsection{Is a data duration of 50~days enough to reliably detect the presence of modes?}
The statistical tests considered in Sect. \ref{ss:detection-prob} suggested that 50~days of data were sufficient to reliably detect the presence of the oscillation power based on the height-to-background ratio. We can  now see if that is true with the algorithms used.
As we used only stars for which we had firm detections of oscillations in the 600-day dataset, we expected to have results for each of the 50-day runs, i.e. 12 results per star. For runs of duration 200 days we expect to have 3 returns etc.
This is not the case as can be seen from Fig.~\ref{returnhisto} where for each of the different methods we plot the fraction of returns for the different data durations as a function of $\Delta \nu$ on a logarithmic scale. The data have been binned for this graph. In general the bin width used is just under 1~$\mu$Hz but bins are combined at high frequencies to improve the statistics where there are few stars in the original sample as can be seen in Fig.~\ref{dnu-distribution}.
As expected, as the run duration increases the general efficacy of each method improves. The exception to this is for the CAN method where the data from the long runs are used to constrain the fitted parameter ranges in the short runs and the method is not `blind'.  The results are summarised in Table~\ref{tabfracreturns}.
Although all methods have difficulties at low frequency, the different methods are clearly somewhat different in the spectral regions where their response is  reliable. Additionally, COR is less effective for mid-range frequencies. The detection capabilities of the EACF method underlie the two methods COR:UP and COR:EACF employed for the determination of $\meandnu$. For this method  the value of the parameter $A_{\rm max}$ as given in  \citet{mosser2009} is important.
The threshold value set for a detection is 8 for rejecting the H0 hypothesis at the 1\% level.
They have shown that the value of $A_{\rm max}$ improves linearly with the duration of the dataset and so we expect a marked improvement as longer datasets are used. This is indeed the case as shown in Table~\ref{tabfracreturns}.
The peak detection methods underlying CAN does depend on a predefined list of stars, and hence this shows the distribution of the type of stars on the list used for the analysis presented here.
The OCT method has issues at the very high frequencies.

\begin{table}
\begin{minipage}{\linewidth}
\caption{Fraction of runs per star, for which results have been returned for $\meandnu$ as a function of timespan of the data, where 12, 6, 3, 1 and 1 runs are available for data of 50, 100, 200, 400 and 600 days length, respectively.}
\label{tabfracreturns}
\centering
\begin{tabular}{lccccc}
\hline\hline
method & 50 & 100 & 200 & 400 & 600  \\
  & days & days & days & days & days \\
\hline
COR:UP & 0.62 & 0.86 & 0.97 & 0.99 & 1 \\
COR:EACF & 0.62 & 0.86 & 0.97 & 0.99 & 1\\
OCT:PS$\otimes$PS & 0.85 & & 0.94 & 0.96 & 1\\
OCT:PS$\otimes$PS (bayesian) & 0.55 & & 0.89 & 0.96 & 0.99\\
\hline
\end{tabular}
\end{minipage}
\end{table}

 \begin{figure*}
\begin{minipage}{0.4\linewidth}
\centering
\includegraphics[width=\linewidth]{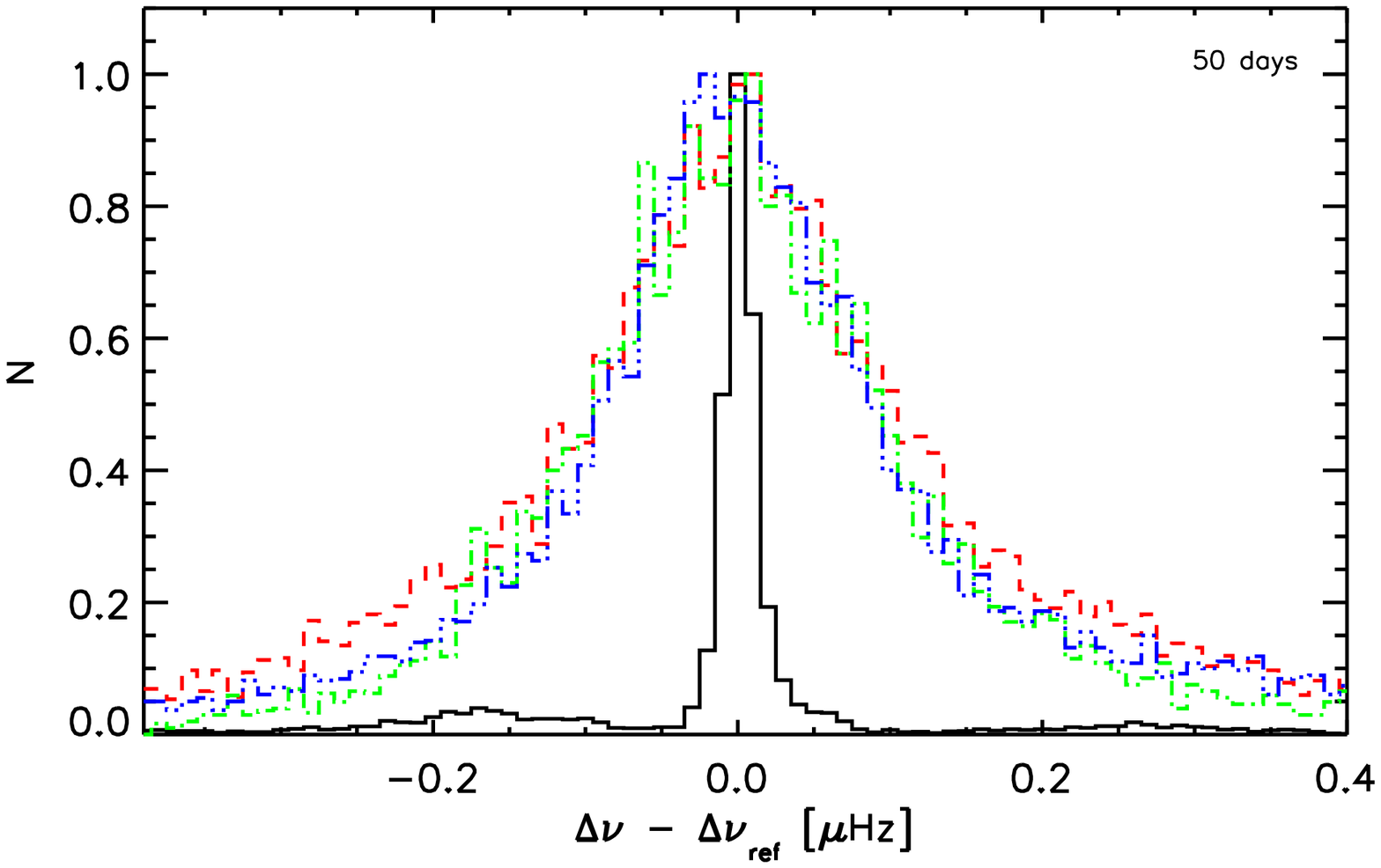}
\end{minipage}
\begin{minipage}{0.2\linewidth}
\centering
\includegraphics[width=\linewidth]{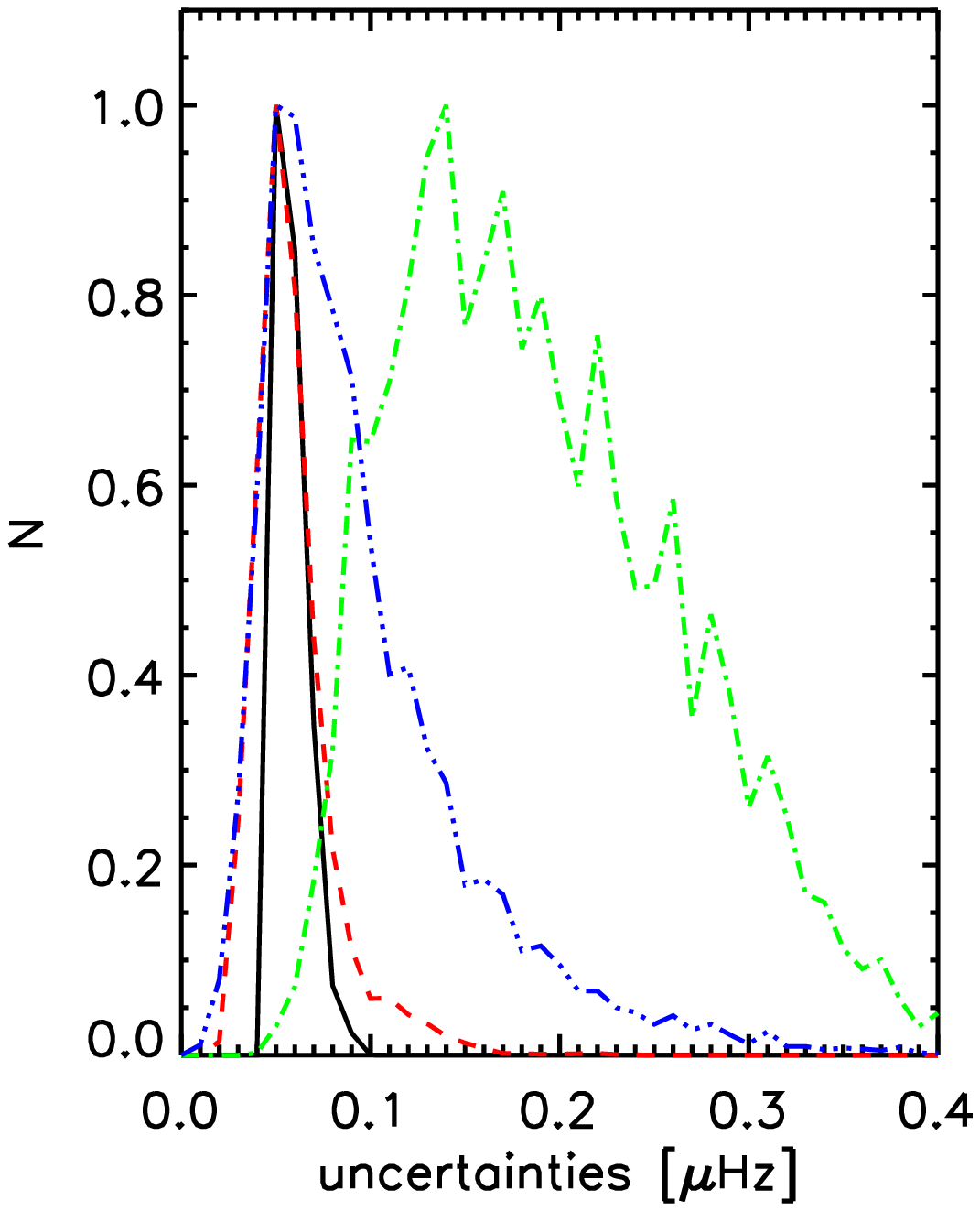}
\end{minipage}
\begin{minipage}{0.4\linewidth}
\centering
\includegraphics[width=\linewidth]{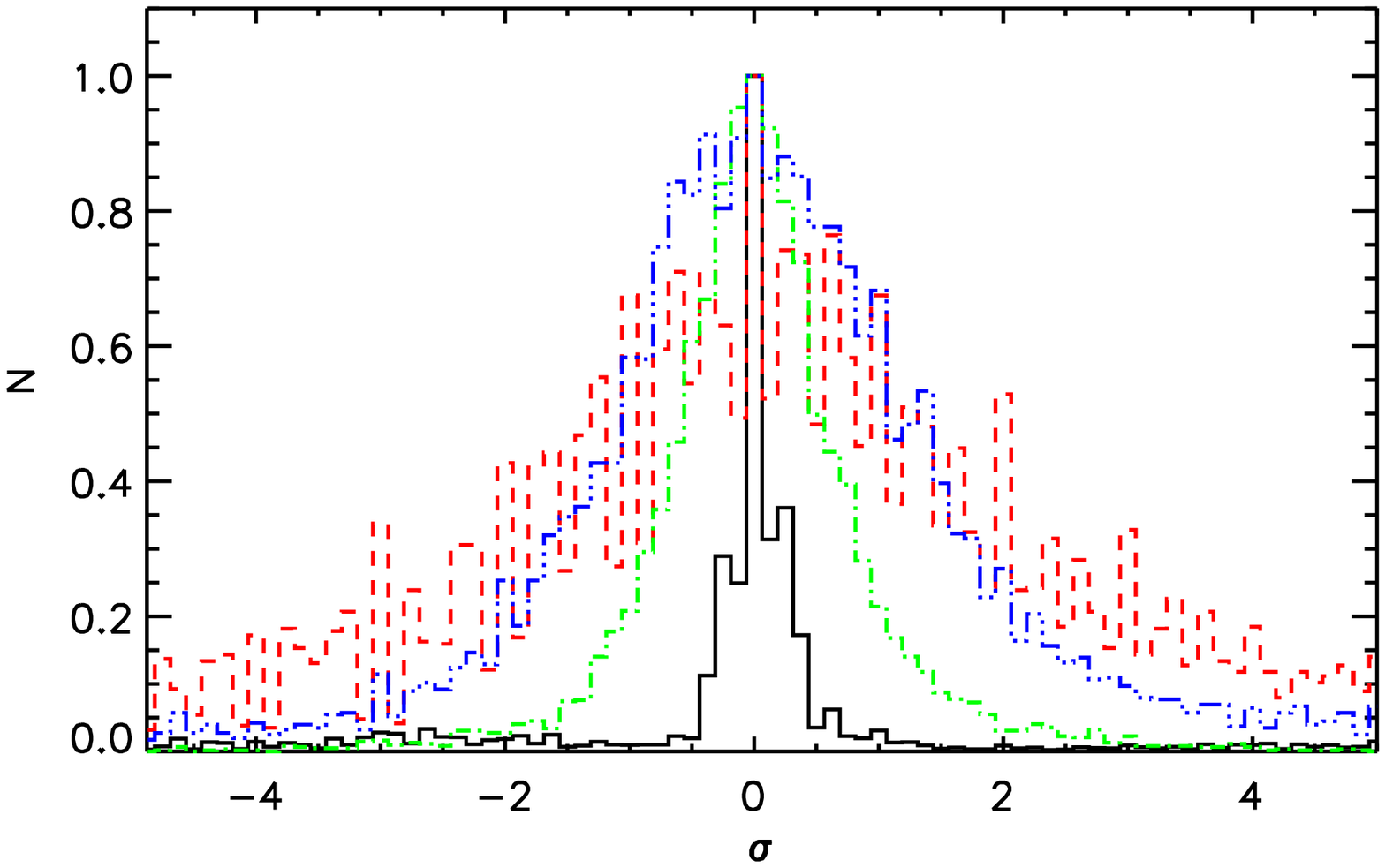}
\end{minipage}
\begin{minipage}{0.4\linewidth}
\centering
\includegraphics[width=\linewidth]{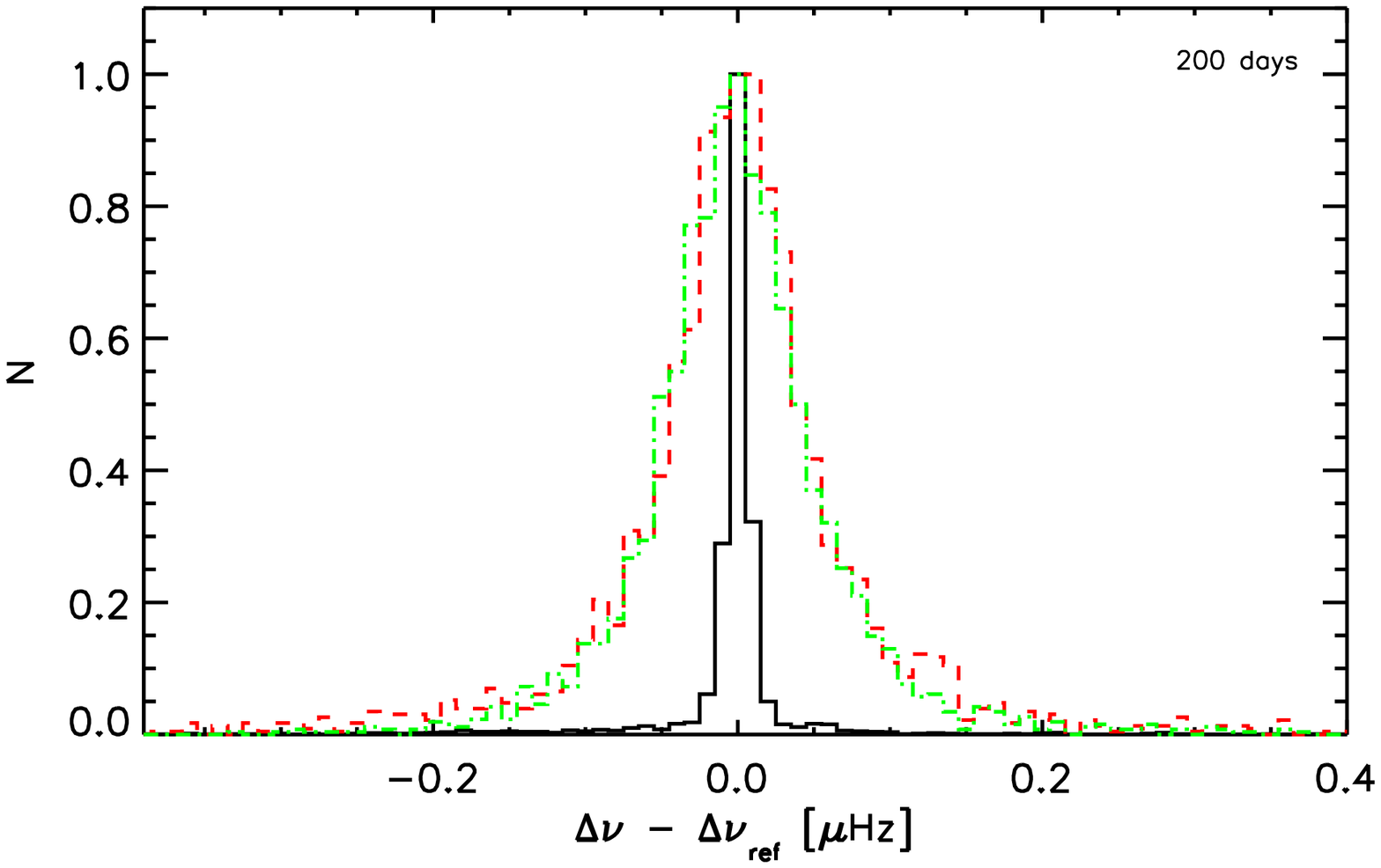}
\end{minipage}
\begin{minipage}{0.2\linewidth}
\centering
\includegraphics[width=\linewidth]{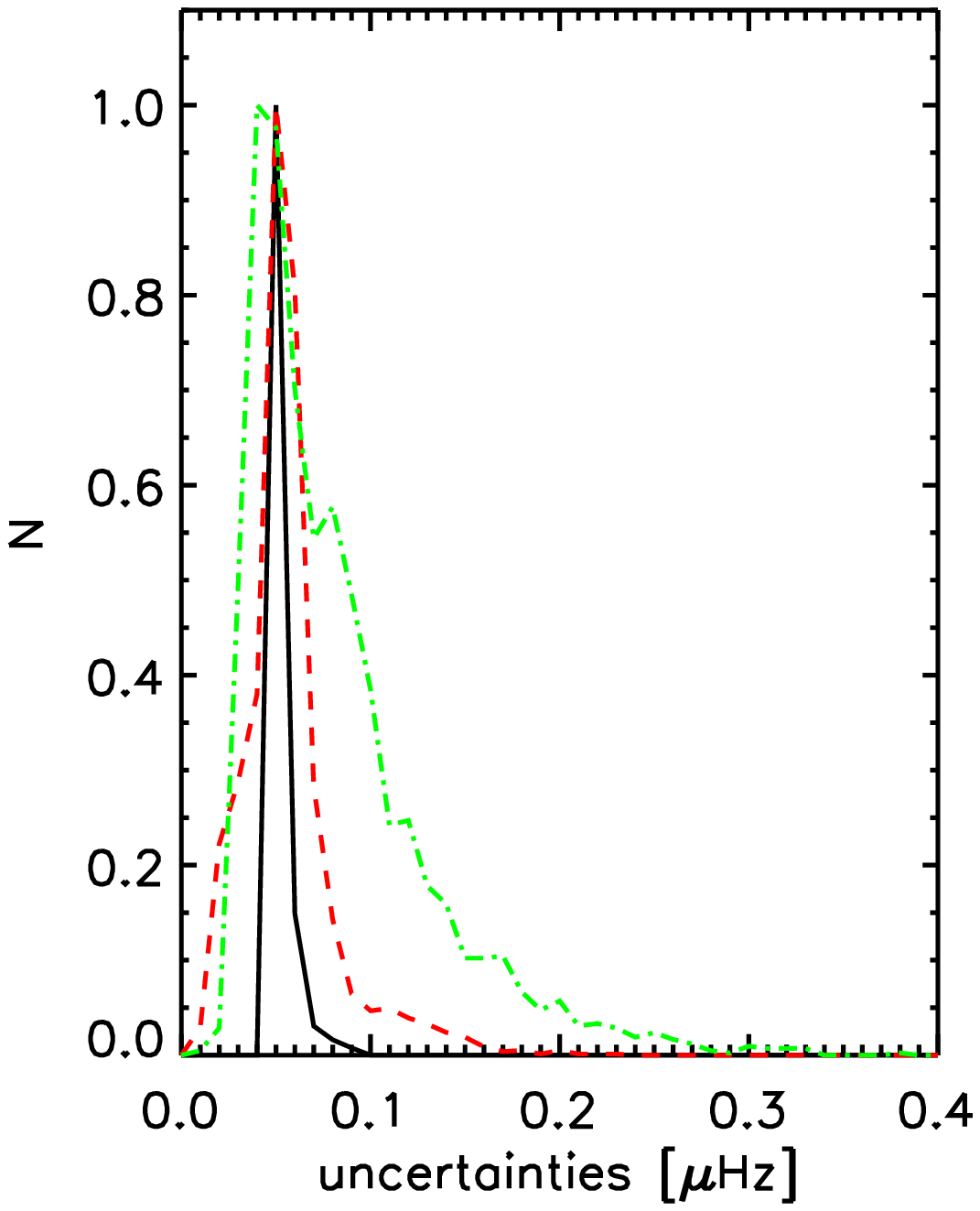}
\end{minipage}
\begin{minipage}{0.4\linewidth}
\centering
\includegraphics[width=\linewidth]{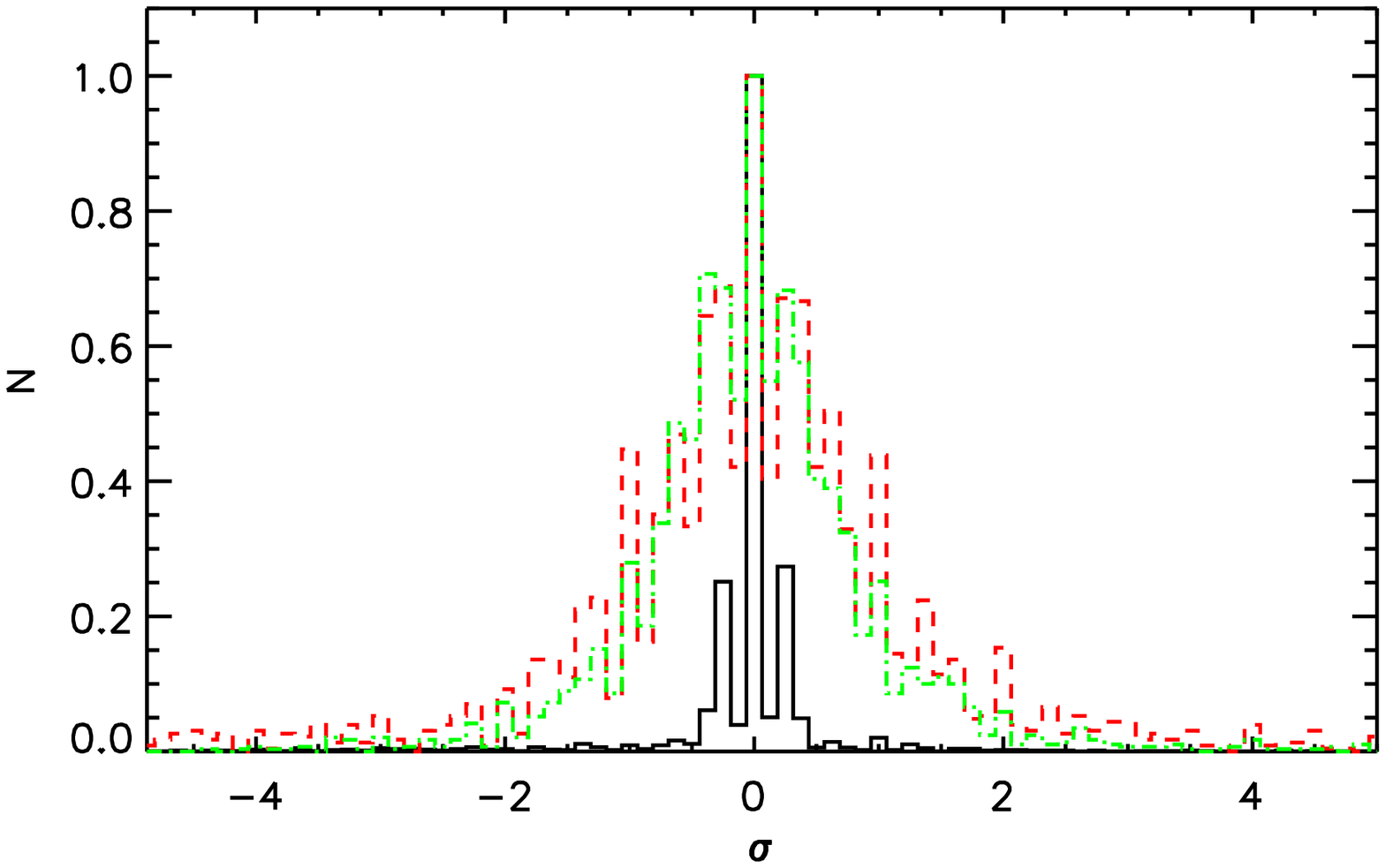}
\end{minipage}
\begin{minipage}{0.4\linewidth}
\centering
\includegraphics[width=\linewidth]{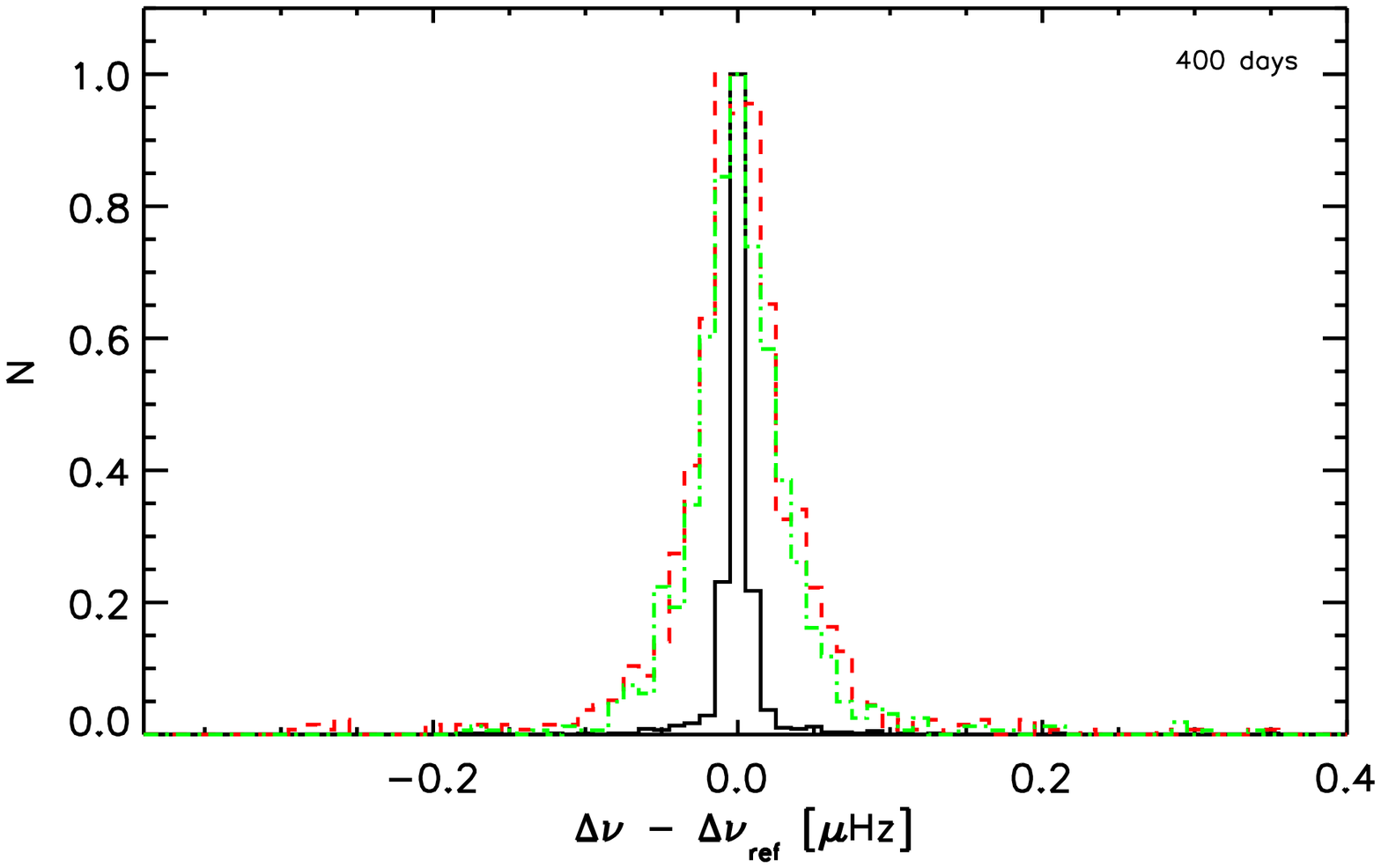}
\end{minipage}
\begin{minipage}{0.2\linewidth}
\centering
\includegraphics[width=\linewidth]{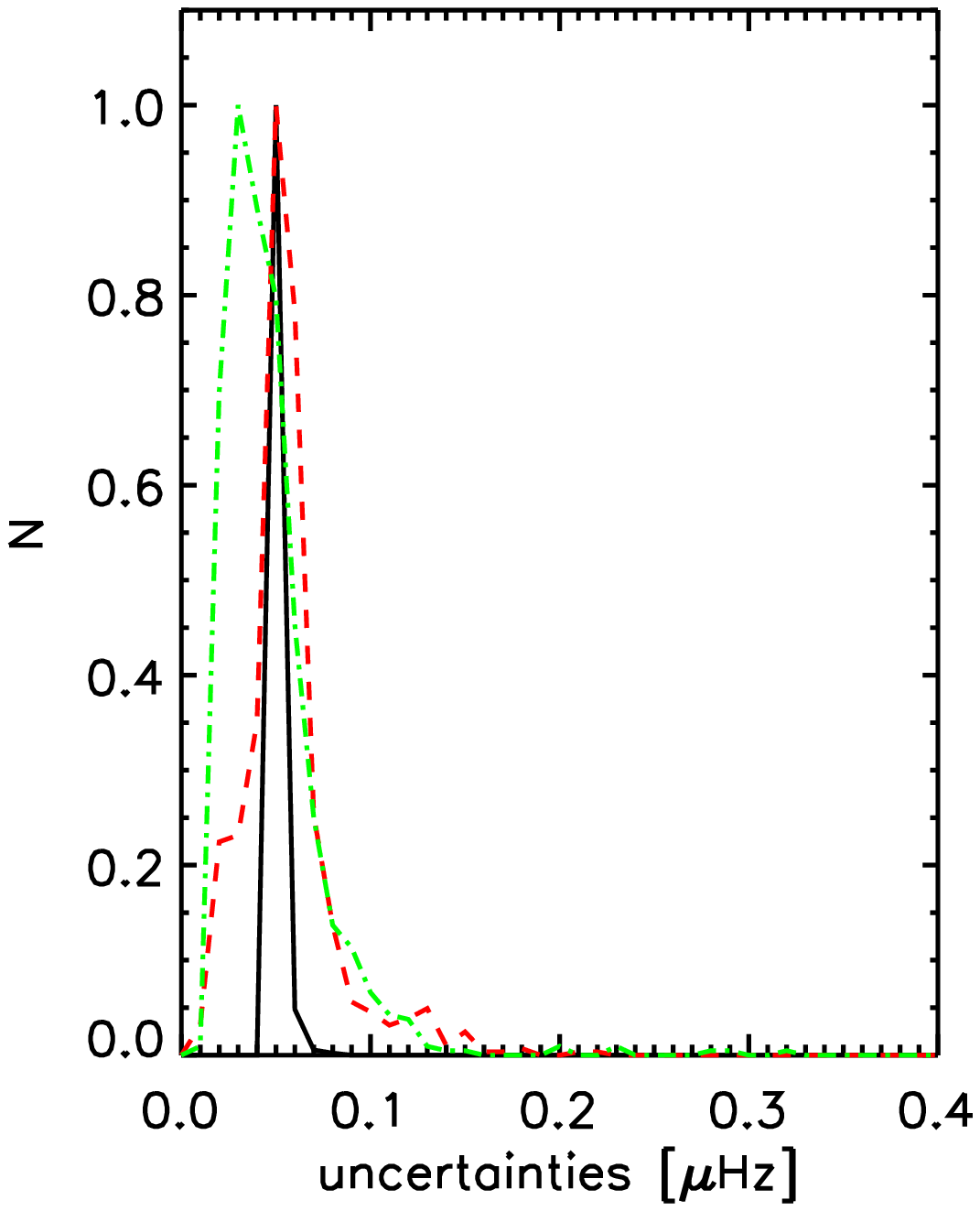}
\end{minipage}
\begin{minipage}{0.4\linewidth}
\centering
\includegraphics[width=\linewidth]{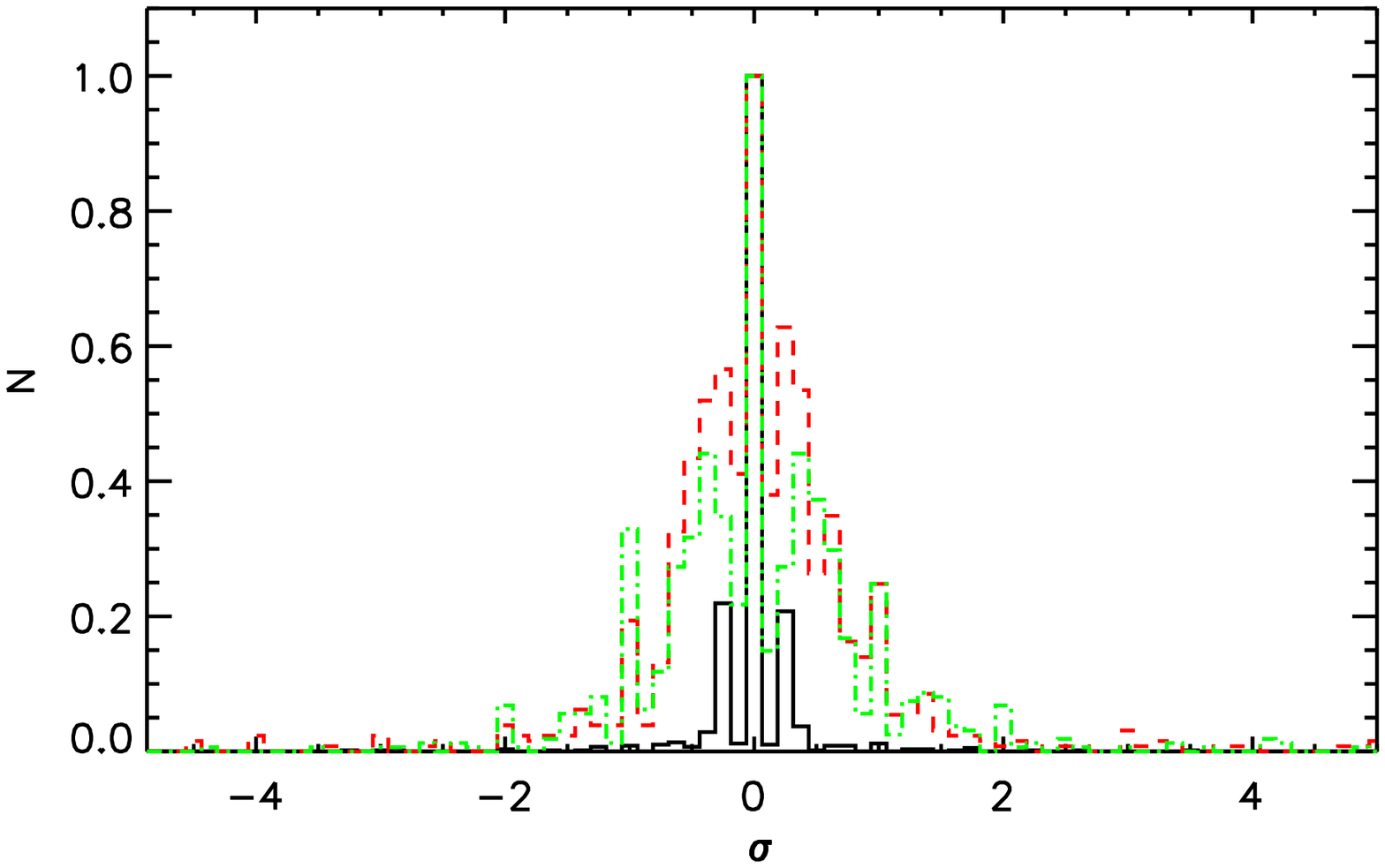}
\end{minipage}
\caption{Normalised distribution of the offset of the individual results from the reference value, i.e., the result of the 600 day run of the same method for the same star (left), and normalised distributions of the uncertainties (centre) for $\meandnu$ for 50 (top), 200 (middle) and 400 day (bottom) datasets. COR:UP, COR:EACF,  CAN and OCT results are indicated in black solid lines, green dashed-dotted lines, blue dashed-triple dotted lines and red dashed lines respectively. The right column shows the normalised distribution of the offset of the individual results divided by its stated uncertainty for data of 50 (top), 200 (middle) and 400 days (bottom) length. Colours and linestyles are the same as in the left panels.}
\label{offsetdnu}
\end{figure*}

 \begin{figure*}
\begin{minipage}{0.4\linewidth}
\centering
\includegraphics[width=\linewidth]{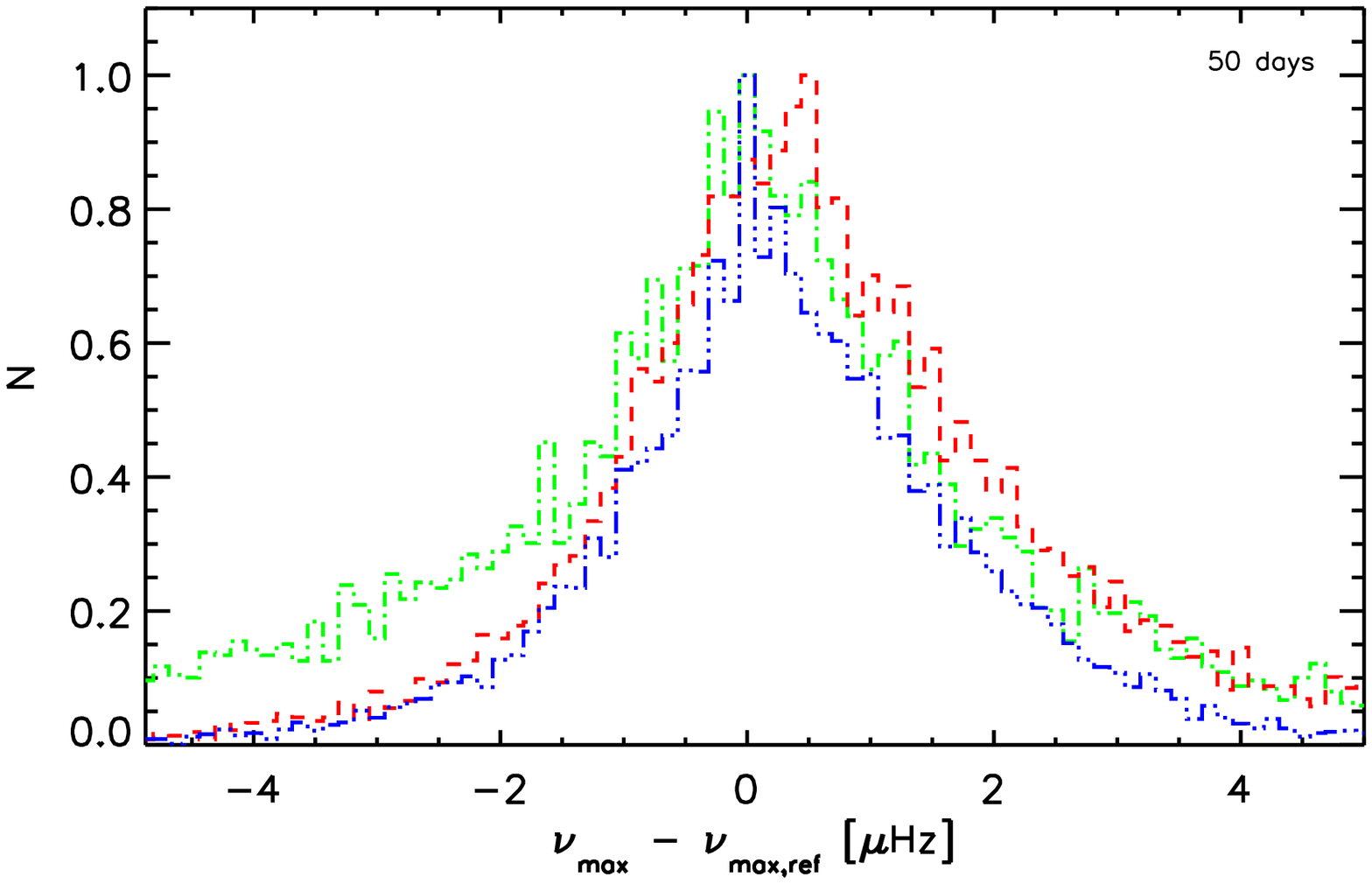}
\end{minipage}
\begin{minipage}{0.2\linewidth}
\centering
\includegraphics[width=\linewidth]{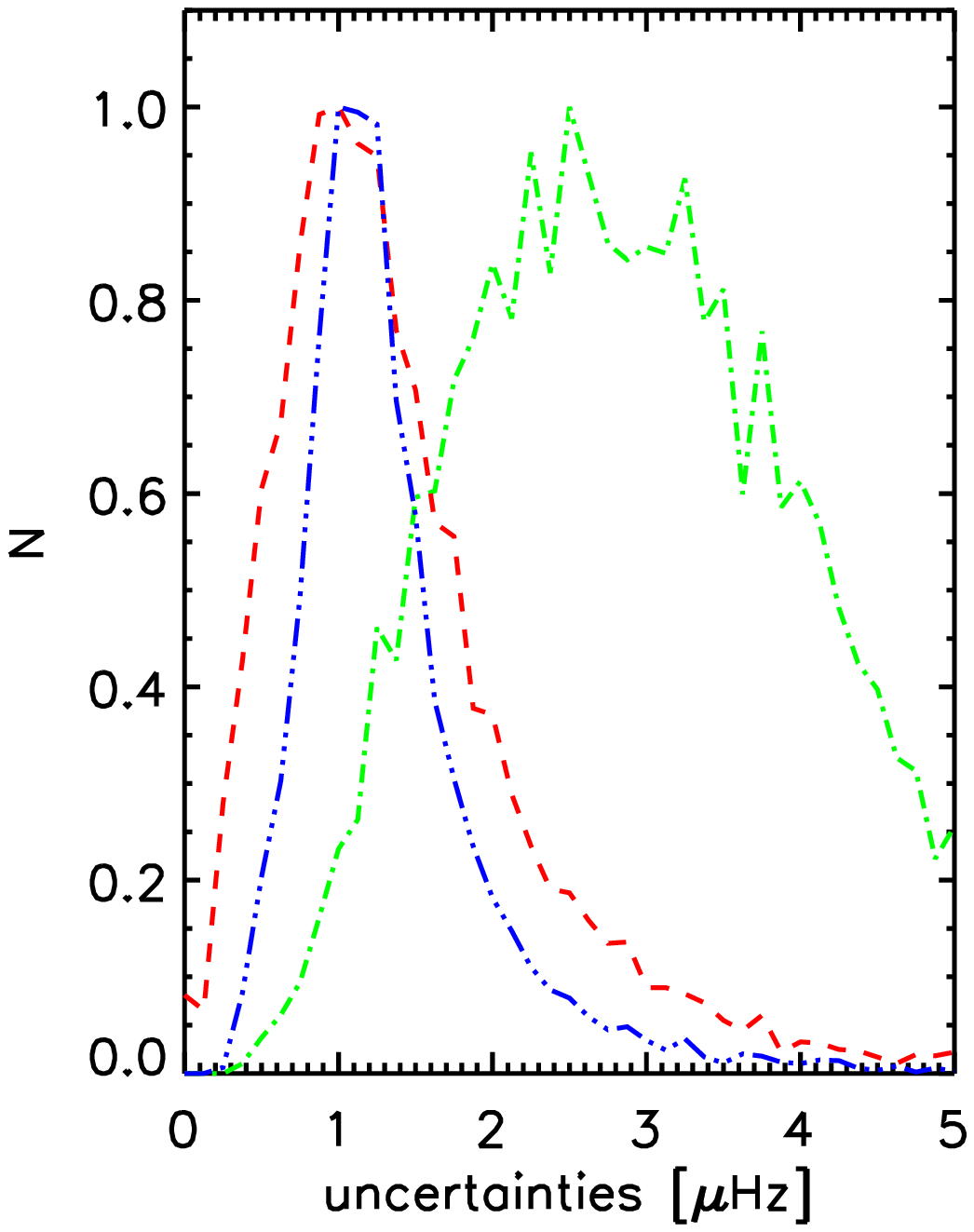}
\end{minipage}
\begin{minipage}{0.4\linewidth}
\centering
\includegraphics[width=\linewidth]{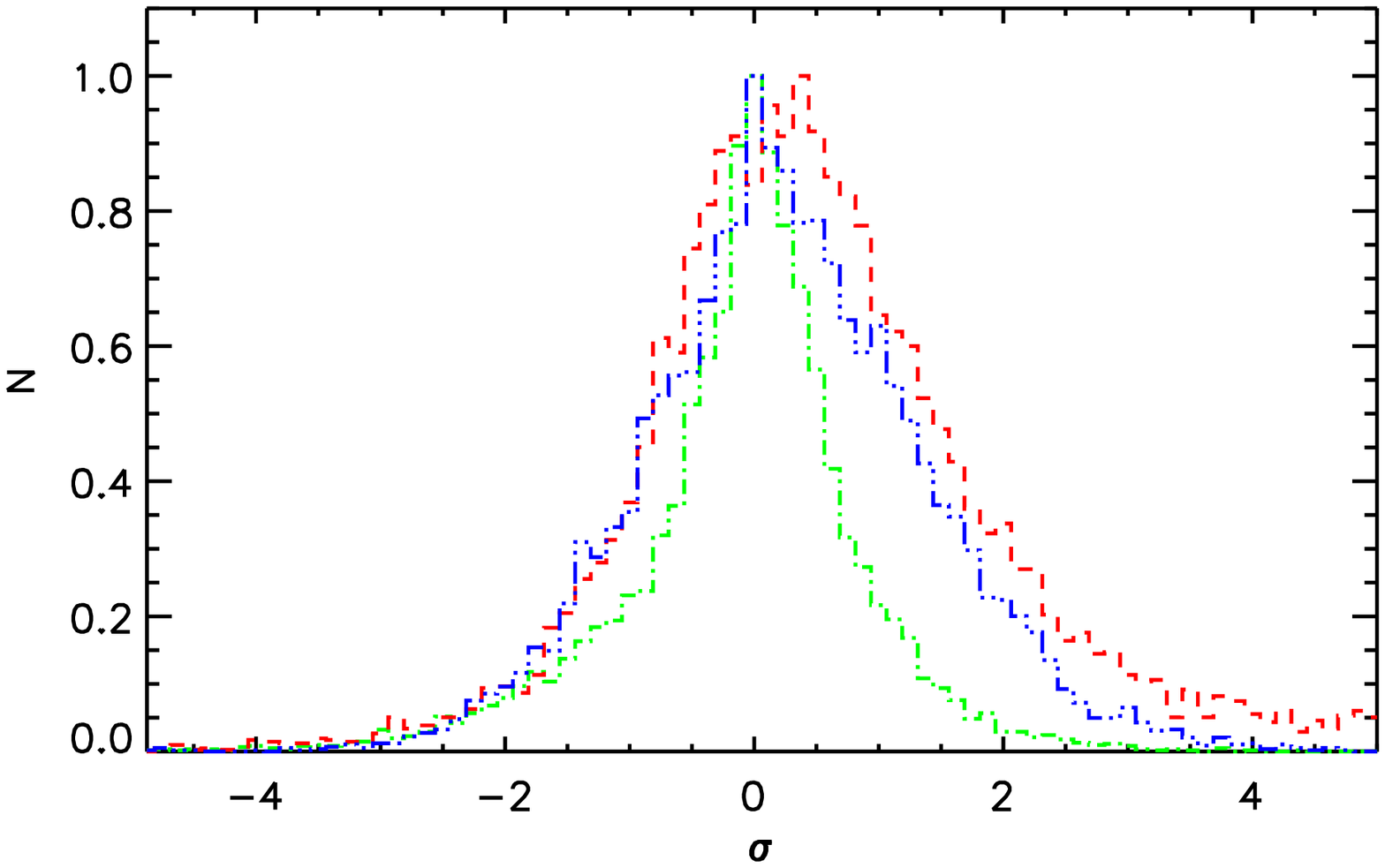}
\end{minipage}
\begin{minipage}{0.4\linewidth}
\centering
\includegraphics[width=\linewidth]{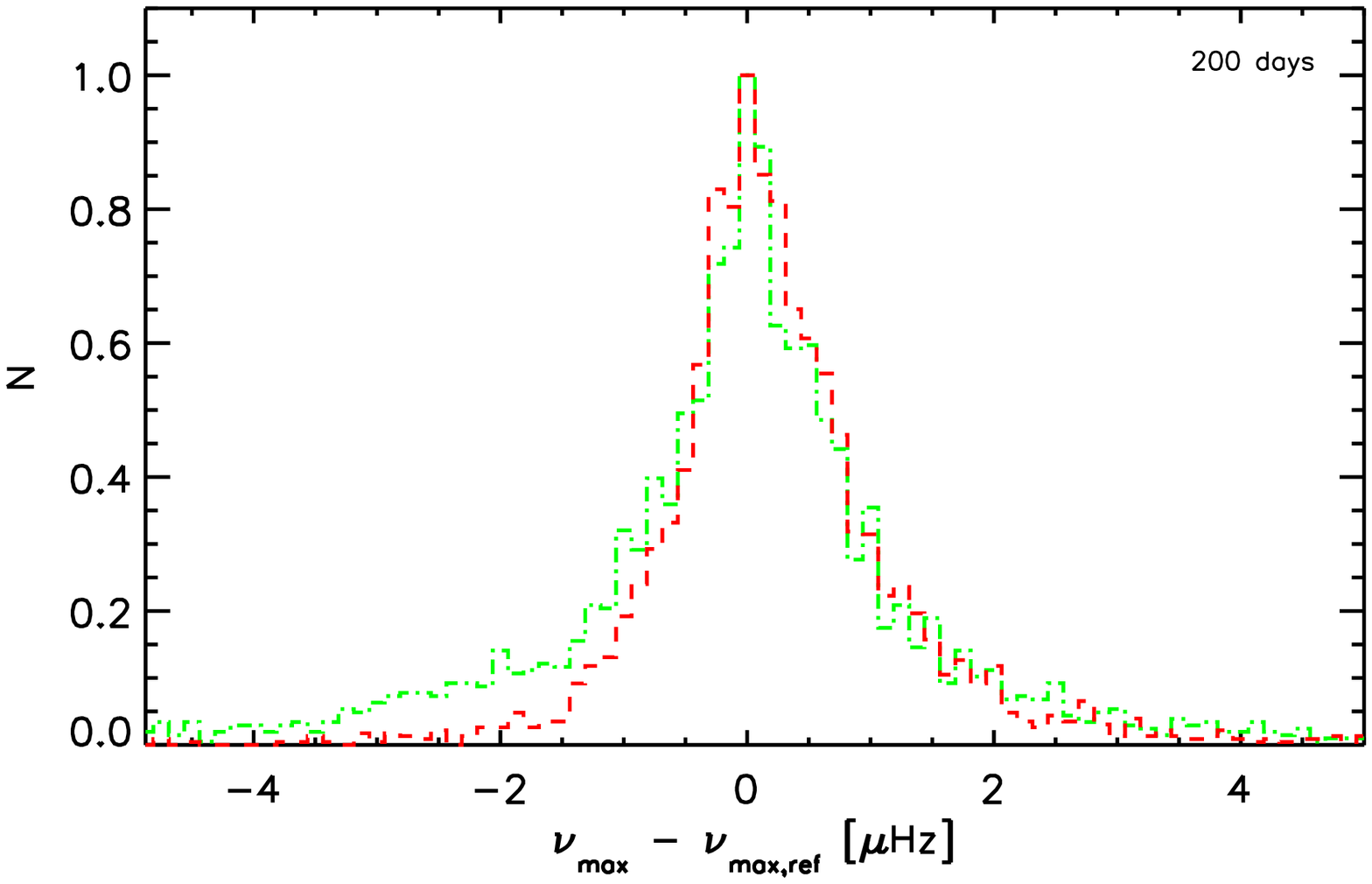}
\end{minipage}
\begin{minipage}{0.2\linewidth}
\centering
\includegraphics[width=\linewidth]{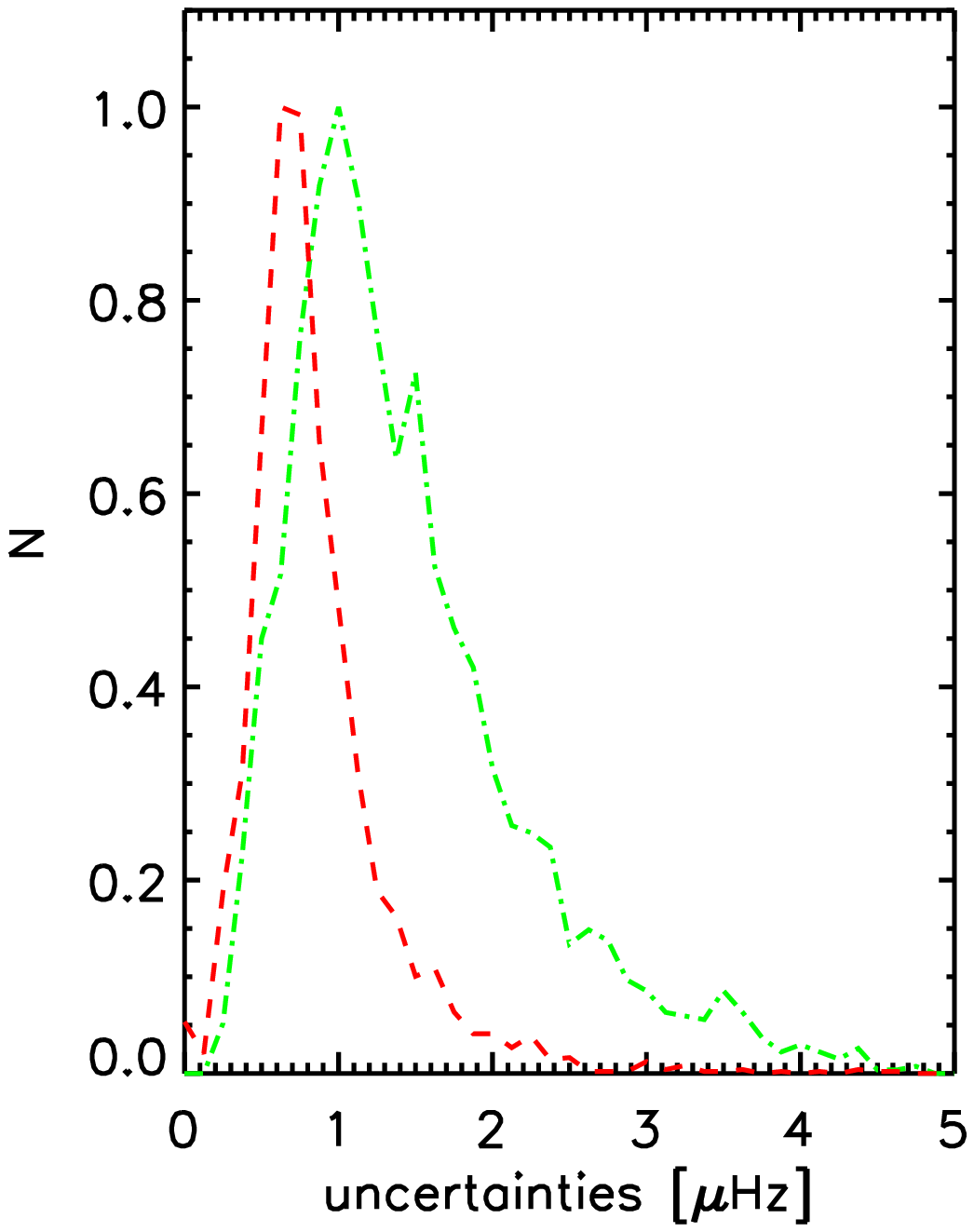}
\end{minipage}
\begin{minipage}{0.4\linewidth}
\centering
\includegraphics[width=\linewidth]{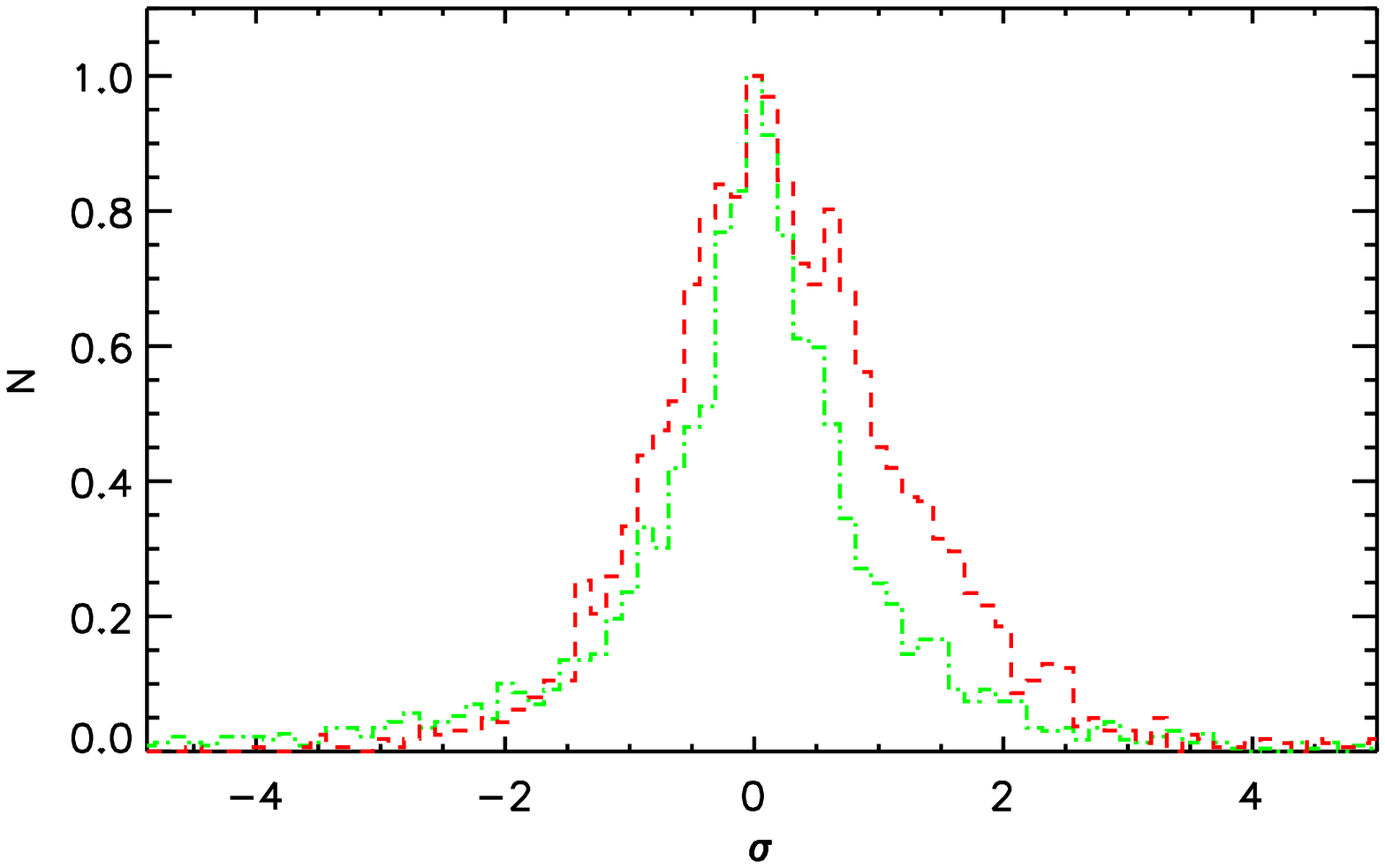}
\end{minipage}
\begin{minipage}{0.4\linewidth}
\centering
\includegraphics[width=\linewidth]{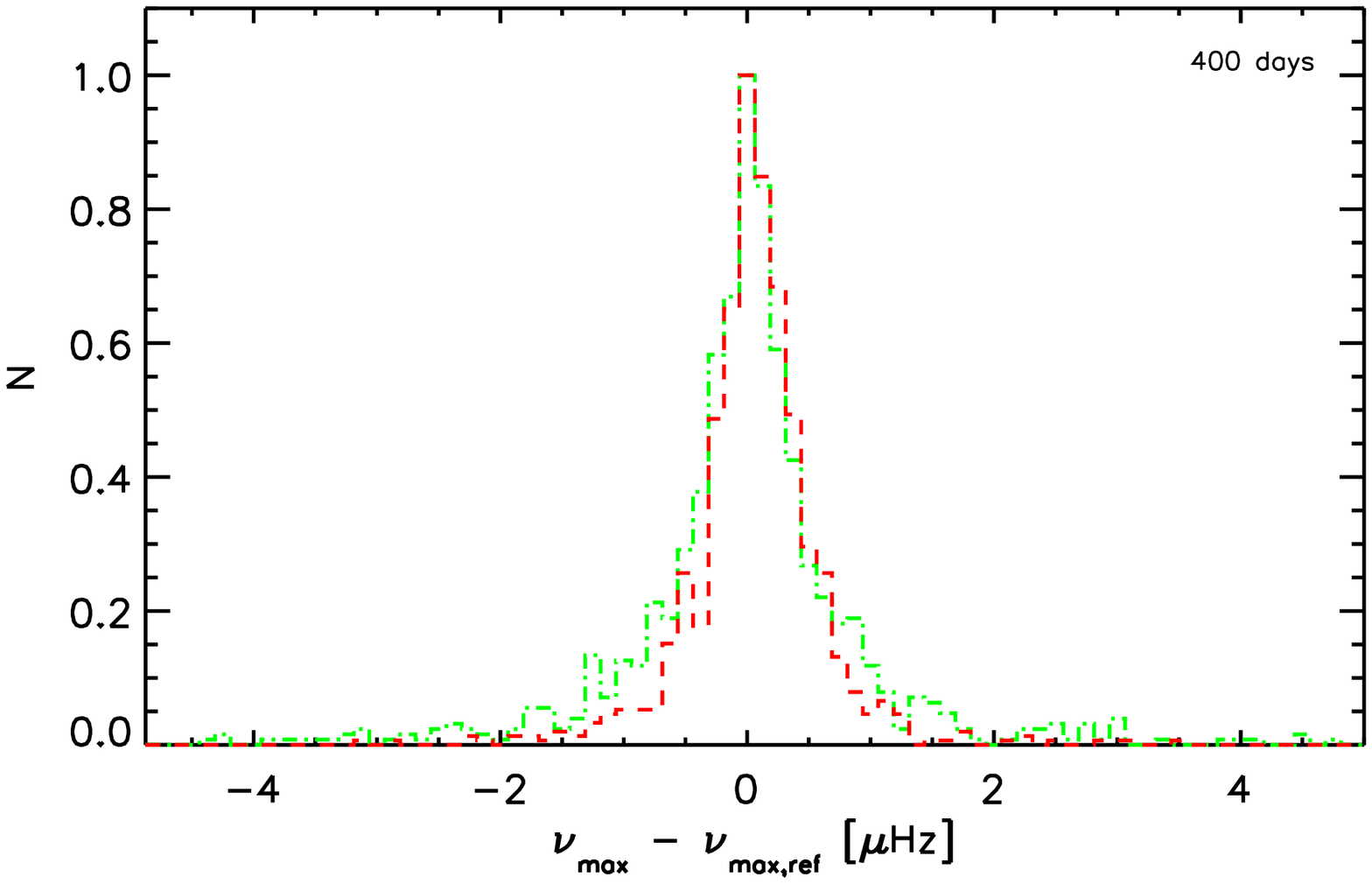}
\end{minipage}
\begin{minipage}{0.2\linewidth}
\centering
\includegraphics[width=\linewidth]{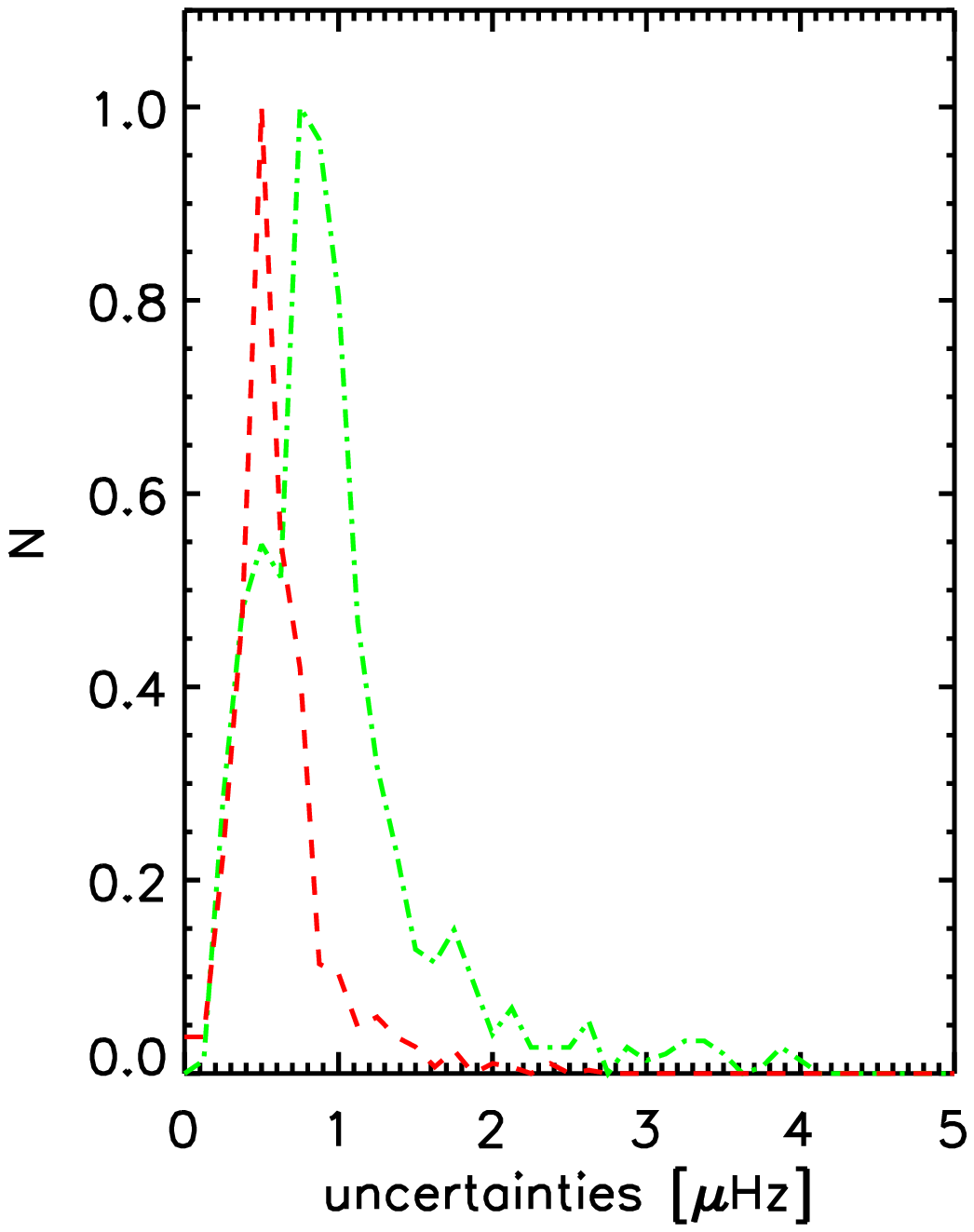}
\end{minipage}
\begin{minipage}{0.4\linewidth}
\centering
\includegraphics[width=\linewidth]{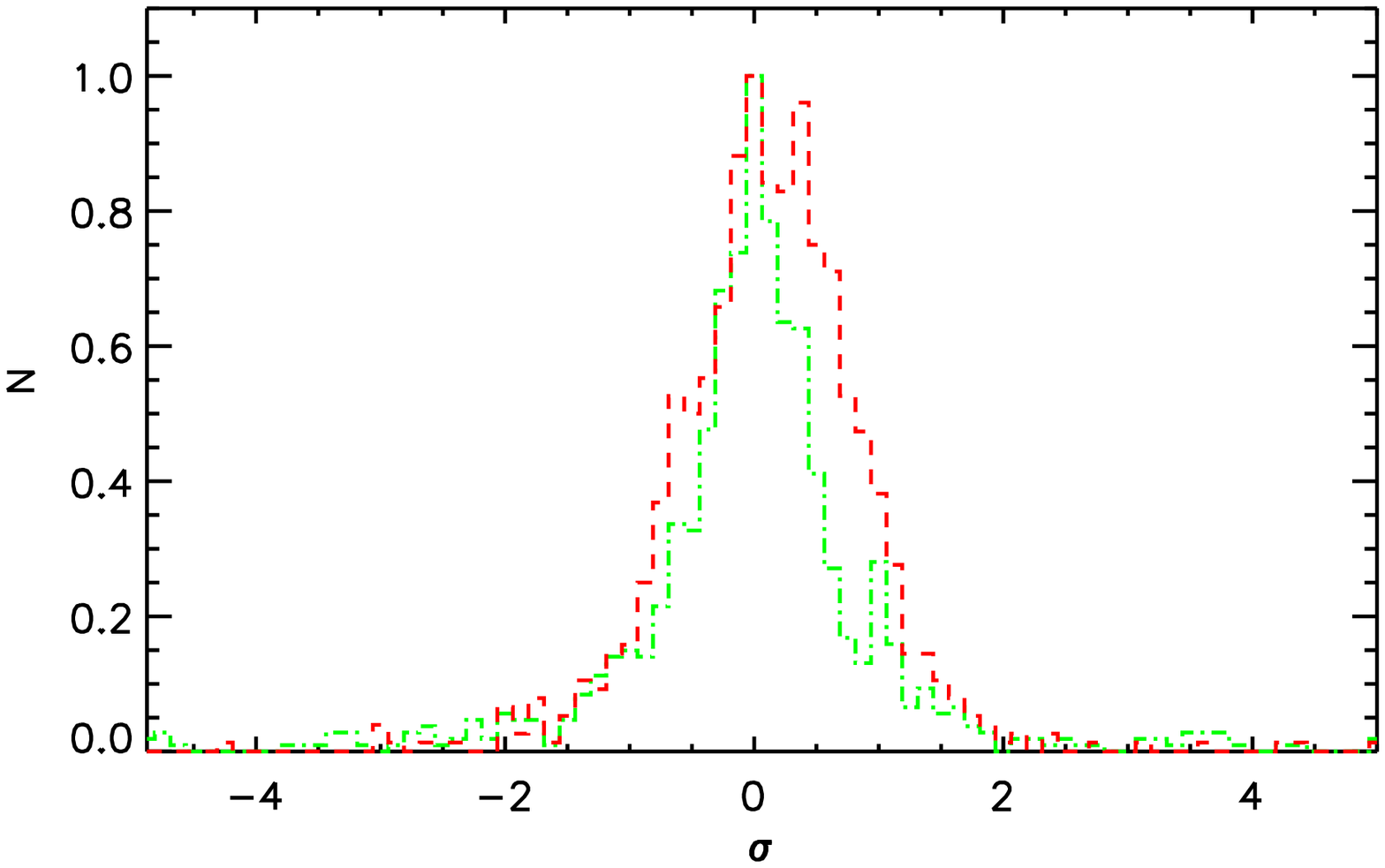}
\end{minipage}
\caption{Same as Fig.~\ref{offsetdnu}, but now for $\nu_{\rm max}$.}
\label{offsetnumax}
\end{figure*}

\begin{figure}
\begin{minipage}{\linewidth}
\centering
\includegraphics[width=\linewidth]{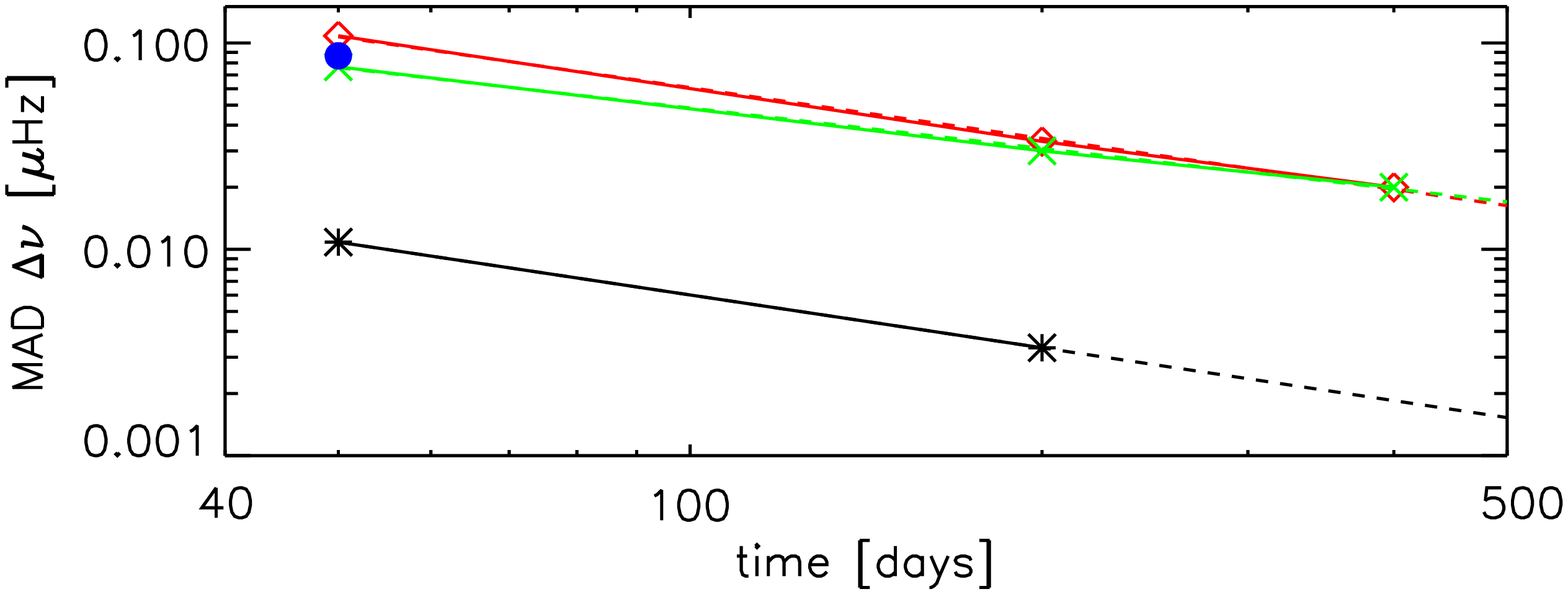}
\end{minipage}
\begin{minipage}{\linewidth}
\centering
\includegraphics[width=\linewidth]{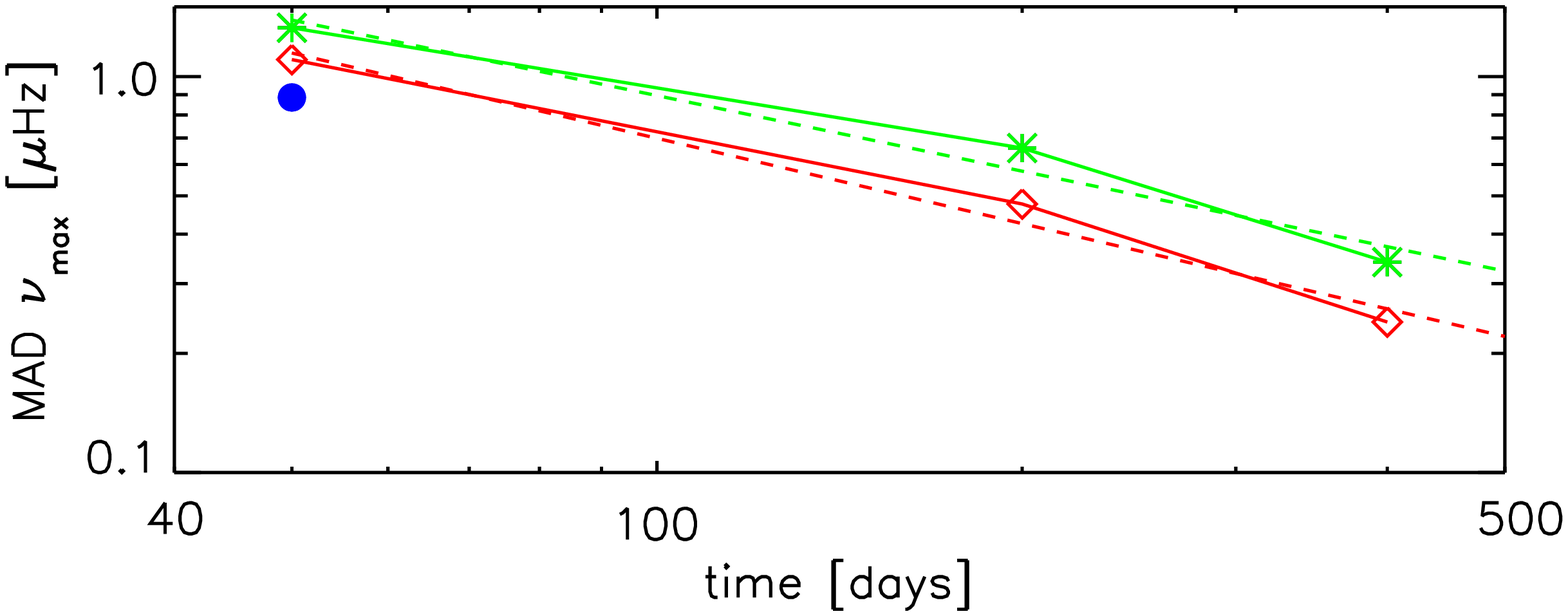}
\end{minipage}
\caption{Median absolute deviations (MAD) observed for $\meandnu$ (top) and $\nu_{\rm max}$ (bottom) for the different methods as a function of the timespan of the dataset. Colour coding the same as in Fig.~\ref{offsetdnu} with red for OCT, green for COR:EACF, black for COR:UP and blue for CAN. The 400 day results of COR:UP agree with the 600 day results and hence the MAD is 0.000 and not shown. The dashed lines indicate linear fits through the data (with same colour-coding) in log-scale. See text for further details.}
\label{rms}
\end{figure}

 \begin{figure*}
\begin{minipage}{0.5\linewidth}
\centering
\includegraphics[width=\linewidth]{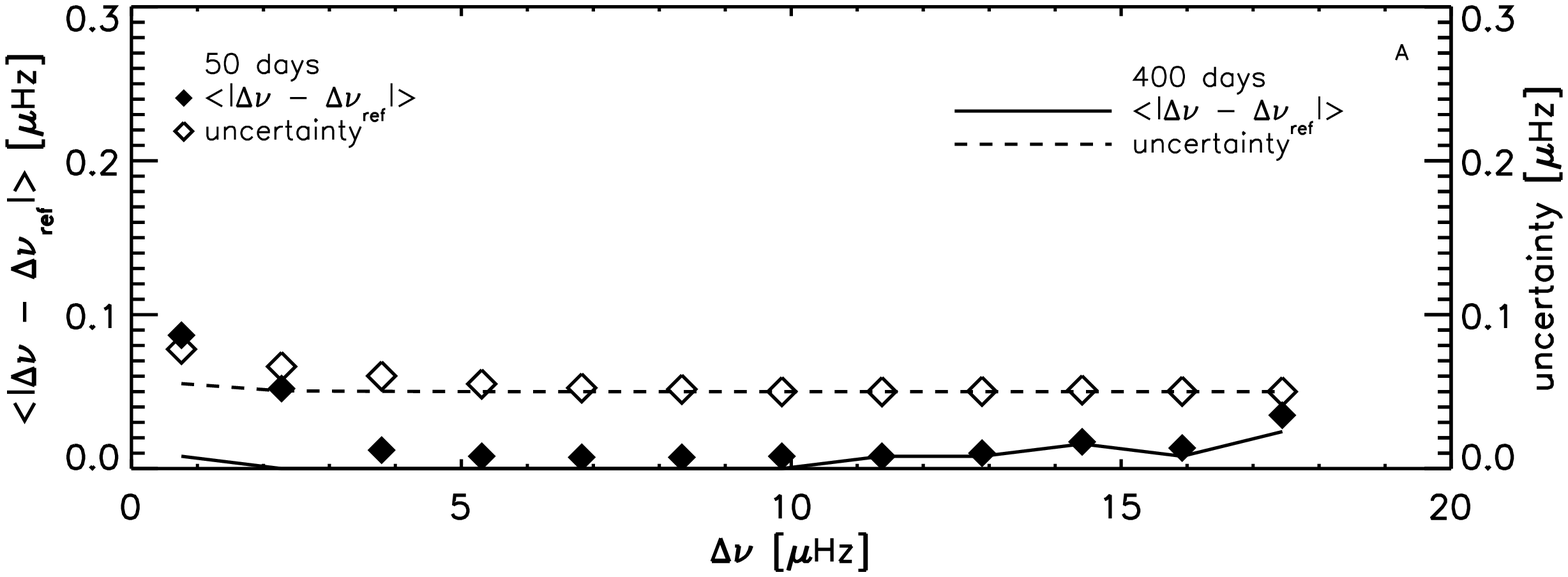}
\end{minipage}
\begin{minipage}{0.5\linewidth}
\centering
\includegraphics[width=0.5\linewidth]{empty.eps}
\end{minipage}
\begin{minipage}{0.5\linewidth}
\centering
\includegraphics[width=\linewidth]{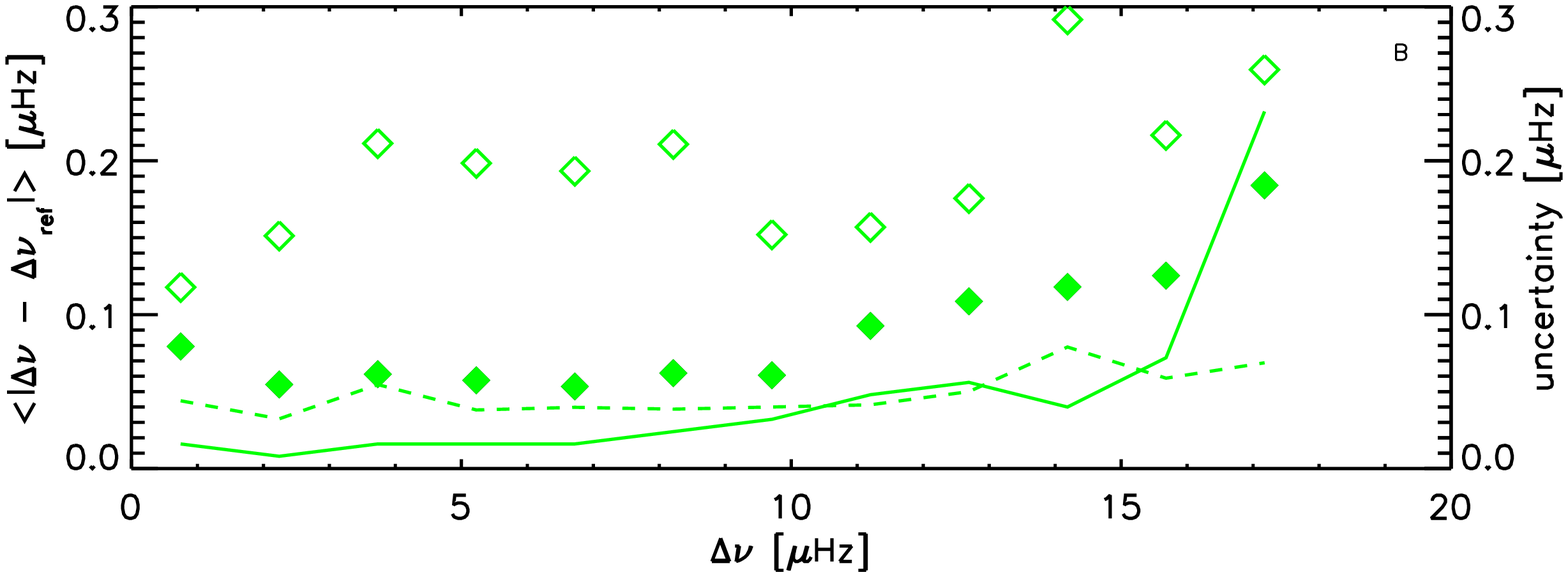}
\end{minipage}
\begin{minipage}{0.5\linewidth}
\centering
\includegraphics[width=\linewidth]{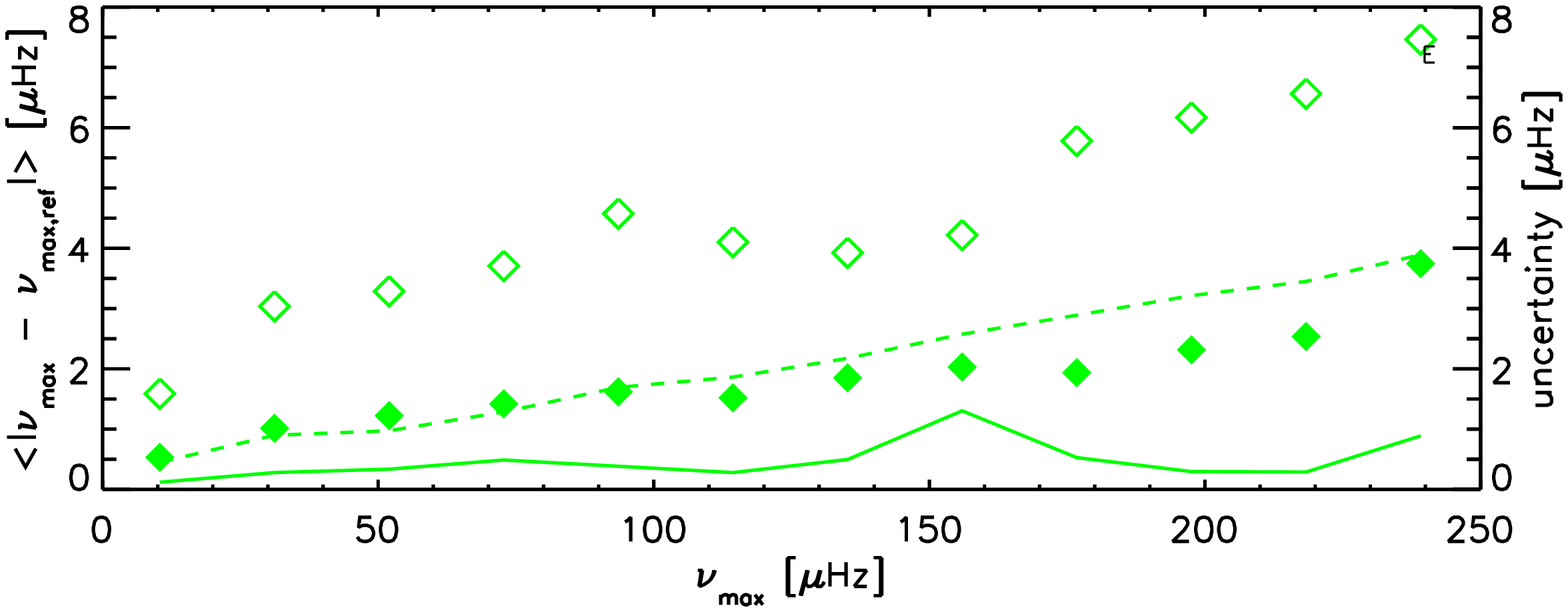}
\end{minipage}
\begin{minipage}{0.5\linewidth}
\centering
\includegraphics[width=\linewidth]{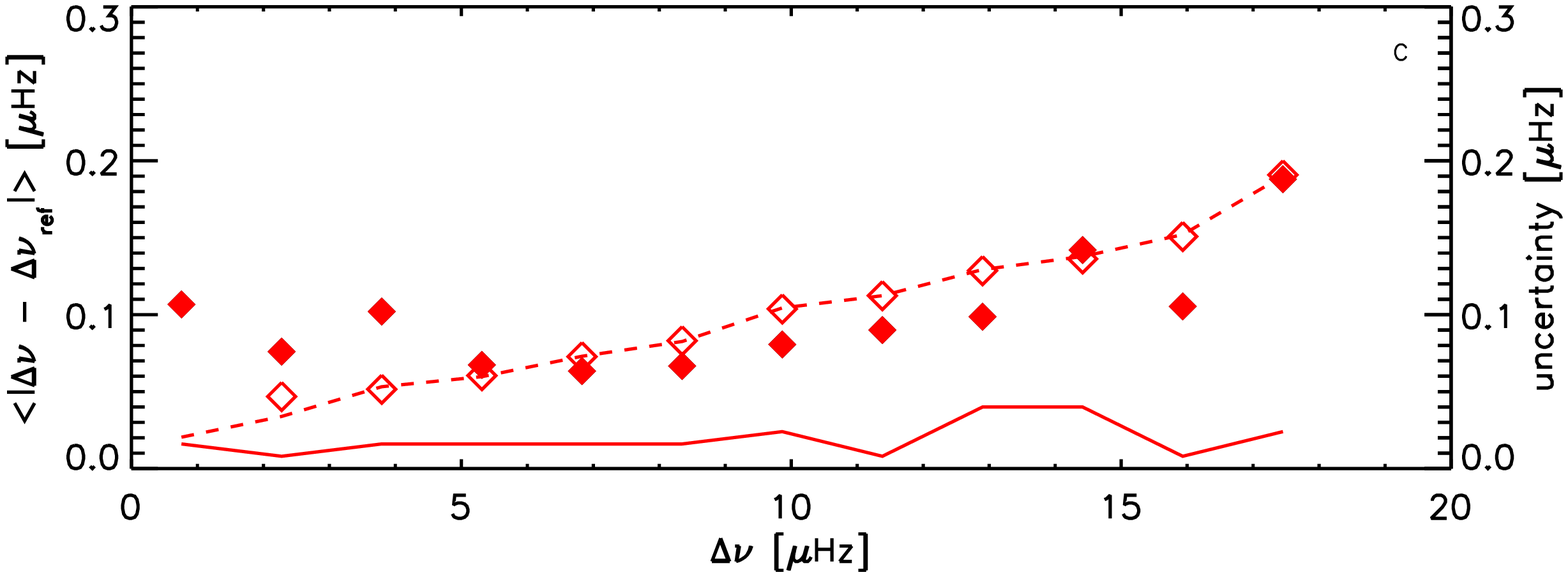}
\end{minipage}
\begin{minipage}{0.5\linewidth}
\centering
\includegraphics[width=\linewidth]{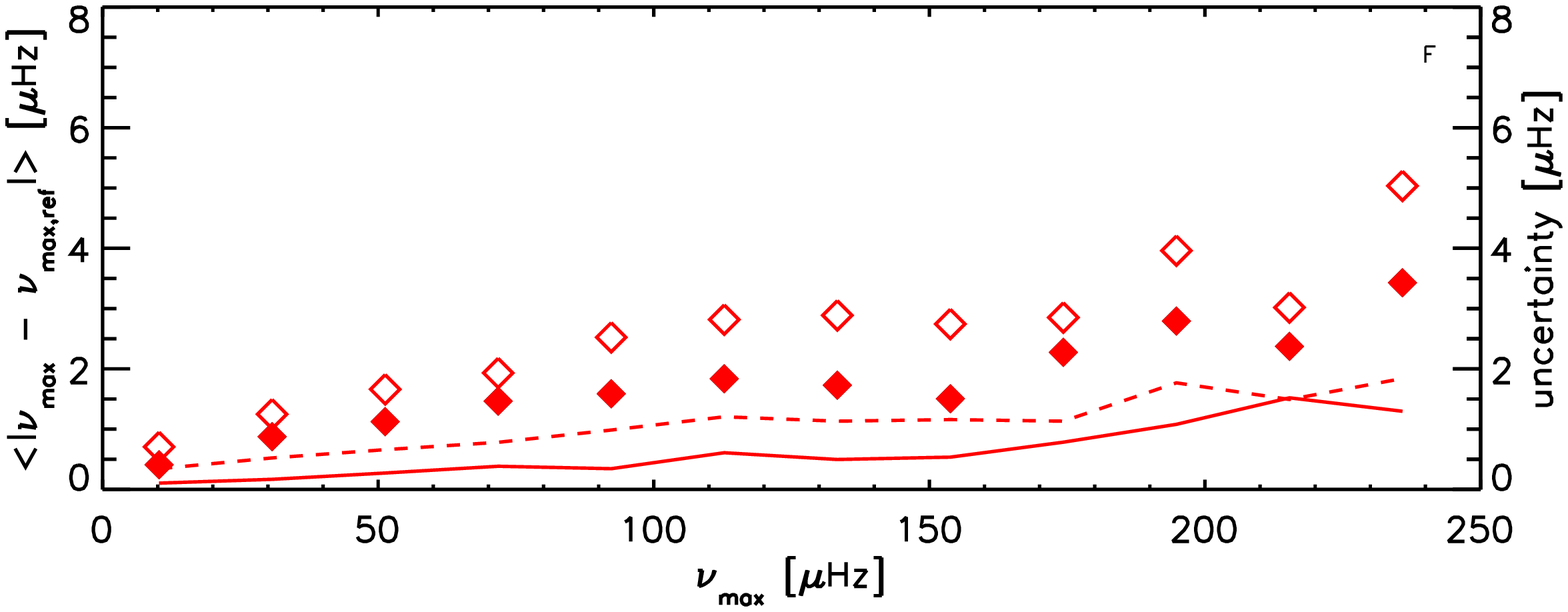}
\end{minipage}
\begin{minipage}{0.5\linewidth}
\centering
\includegraphics[width=\linewidth]{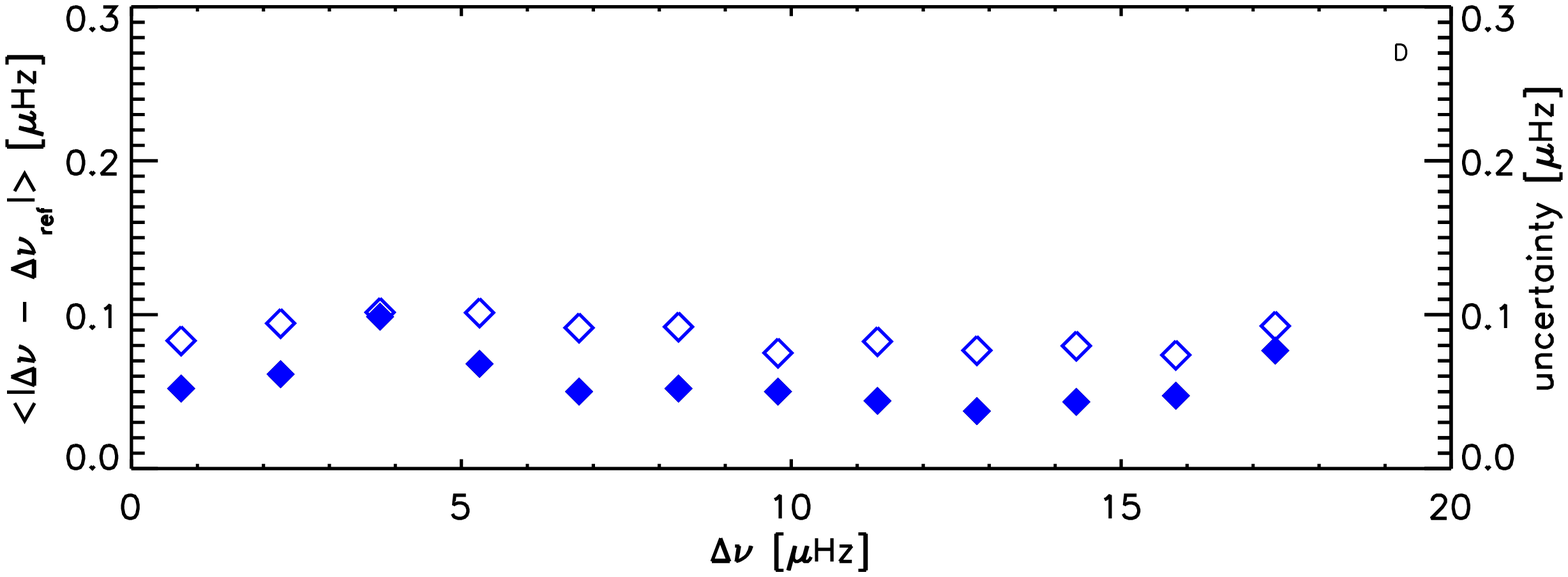}
\end{minipage}
\begin{minipage}{0.5\linewidth}
\centering
\includegraphics[width=\linewidth]{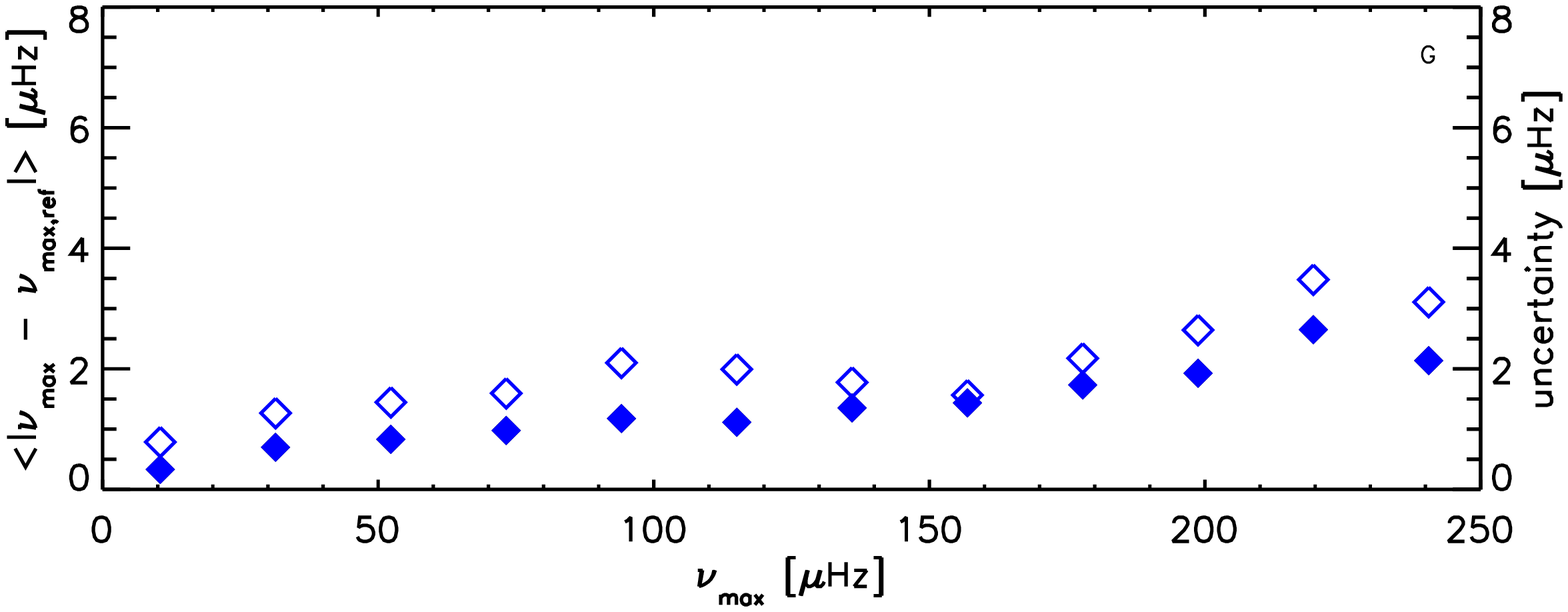}
\end{minipage}
\caption{Left:  Uncertainties (open diamonds: 50 days, dashed line: 400 days) and mean absolute deviations multiplied by 0.8 (see text, filled diamonds: 50 days, solid line: 400 days) in $\meandnu$ as a function of $\meandnu$ for results from the different methods: COR:UP (panel A), COR:EACF (panel B), OCT:PS$\otimes$PS (panel C) and CAN (panel D). Right: Uncertainties (open diamonds: 50 days, dashed line: 400 days) and mean absolute deviations (filled diamonds: 50 days, solid line: 400 days) in $\nu_{\rm max}$ as a function of $\nu_{\rm max}$ for results from three different methods: COR (panel E), OCT (panel F) and CAN (panel G). }
\label{errdevdnu}
\end{figure*}

It is important to note that all these algorithms rely not on detecting the presence of the oscillation power but instead they look for patterns in the spectrum that are the consequence of the regular spacing in the spectrum of the modes. In looking at the fractions of the stars for which we detect regular mode structure we are really considering a different measure from the $H/B$ ratio-based derivations in Sect. \ref{ss:detection-prob}, hence we are comparing two different strategies.
In a dataset of 50 days the modes are barely resolved \citep{baudin2011,baudincor2011} and so the amplitude of the mode in the spectrum is very variable. In fact the power varies as  $\chi ^2$ with 2dof, which means that the probability distribution of power is negative exponential and it is not unusual for a particular mode to be essentially absent.
As the duration of the dataset increases and the modes become resolved this is less of a problem. From Table~\ref{tabfracreturns} we see that for timeseries of 100 days length we have just about 85\% return and for 200 days long timeseries about 95\%, increasing to over 95\% for 400 day datasets. The OCT:PS$\otimes$PS(bayesian) is most sensitive to the timespan of the data and is only as reliable in detecting the oscillations as the other methods for timeseries of 400 days or longer.
These tests suggest that in short datasets the height-to-background would be a more reliable method to detect oscillations as opposed to the currently developed methods based on the regularity of the frequency pattern.

So the simple answer to the question posed at the beginning of this section is `no, 50 days is not enough to be certain to pick up more than 90\% of the oscillations with the currently employed methods, but with methods based on height-to-background it is predicted that it would be possible to obtain reliable results in such short data-sets.' 

\subsection{Dependence of $\nu_{\rm max}$, $\meandnu$ and their quoted uncertainties on the length of the timeseries}
We have looked at the likelihood of the modes being detected in datasets of differing lengths but there is another important consideration. Here we consider the precision of these results by comparing them with reference values. Because all methods use slightly different definitions for $\nu_{\rm max}$ and $\meandnu$ and we first aim to investigate the influence of the timespan only, we use the results of the 600 day run of a particular method as the reference to compare results of  the shorter runs of that same method with. We evaluate both the deviation of the returned values from the reference values and the quoted uncertainty on the value. 

We first explore how the deviations from the reference value and the uncertainties compare for the different data durations. The left panels of Figs.~\ref{offsetdnu} and \ref{offsetnumax} show the distribution of the deviation of the individual results from their respective reference values for each of the global parameters considered here for data with a timespan of 50, 200 and 400 days for the range of methods employed. The different timespans are shown in different rows and the different methods are plotted in different colours with different line styles.
The left hand panels of Figs.~\ref{offsetdnu} and~\ref{offsetnumax} show that except for the measure of $\meandnu$ by COR:UP the spread in the difference decreases with increasing timespan of the data. The reason for the difference in behaviour of COR:UP originates from the fact that this method applies the additional constraint of a regular pattern on the spectrum. The decrease of the spread with increasing timespan raises the question whether we can expect further improvements from even longer datasets. Therefore, we show the spread as a function of timespan in Fig.~\ref{rms}. The spread for COR:UP is 0.000 at 400 days (not shown) and this method is very reliable at determining the $\meandnu$ even for short datasets. The decreasing trend of the spread in the global oscillation parameters for longer timeseries of the other methods suggests that longer datasets would still improve the precision of the obtained parameters. To investigate this further we show linear fits in log-scale through the MAD values of each method. When extrapolating these fits to 2000 days ($\sim$5.5 yrs, which is the current predicted length of the mission), this would imply a reduction in the MAD of at least a factor of 10 (for $\meandnu$ factors of 23, 11 and 20 for COR:UP, COR:EACF and OCT respectively and for $\nu_{\rm max}$ factors of 10 and 14 for COR:EACF and OCT).
In addition to the spread in the results we also checked for potential biases. It is noticeable that the offsets are not zero even though the method is its own reference. These biases are more clearly visible in the right hand panels where we show the distribution of the offsets divided by the quoted uncertainty ($\sigma$) expressed in dimensionless units.

We now turn to the uncertainties reported by the different methods. The normalised distributions of these uncertainties are shown in the central columns of Figs.~\ref{offsetdnu} and \ref{offsetnumax}.
Again we can see that for some run durations, the different methods produce similar uncertainties and for others they differ.
An important consideration is the validity of the uncertainties as a guide to the reliability of the returned results. To this end, in the right hand column we show the distribution of the offsets divided by their individual quoted uncertainties expressed in dimensionless units. In case of statistically reliable quoted uncertainties we would expect the distributions to have a width of $\pm1\sigma$ at half maximum. In case of a wider distribution the uncertainties are underestimated and a more narrow distribution indicates overestimated uncertainties. For $\meandnu$ we see that OCT and CAN provide realistic uncertainties for runs of 50 day lengths, although the tails of the distribution of OCT are well-populated. Both methods of COR seem to overestimate the uncertainties. The banded nature of the COR:UP results is a byproduct of the method used to find the peak in the autocorrelation function. For longer datasets all methods seem to overestimated the uncertainties to a certain extend. Similar conclusions can be drawn for the results of $\nu_{\rm max}$ in the right hand panels of Fig.~\ref{offsetnumax}.

The measures described above do however average over the frequency range at which the oscillations occur and the uncertainty might be expected to be a function of frequency.  Fig.~\ref{errdevdnu} shows the frequency dependence of the mean uncertainty and the median absolute deviation (MAD) for several methods. For a Gaussian distribution (white noise), the typical ratio of root median square deviation to MAD is roughly 0.8. So we multiply the MAD by 0.8 in order to compare it with the typical uncertainty.
The left hand column is for $\meandnu$ and the right hand column is for $\nu_{\rm max}$.
Each graph in the figure corresponds to a different method and allows us to illustrate how the deviations (MAD) and uncertainties correspond for a given method at the longest and the shortest data duration, i.e., 400 and 50 day long datasets.

It is clear that although there is some consistency in the curves for any one method, the frequency dependencies of the uncertainty and of the deviation are not identical. We now discuss each method in turn starting with $\meandnu$. For COR:UP, we see again that the results for 50 or 400 day long timeseries are remarkably similar. Significant improvement can only be seen at low frequencies. The uncertainties seem to be overestimated. For COR:EACF, at 50 days the uncertainties are over-estimated. However, at 400 days the uncertainties and MAD have reduced and are more closely in agreement except for the highest frequencies where there are not many stars. For OCT:PS$\otimes$PS, at 50 days the uncertainties are underestimated at low and medium frequencies with the agreement steadily improving as the frequency increases. At 400 days, the uncertainties are progressively over estimated. The determination of values as illustrated by the value of MAD improves in the longer datasets. Finally, we consider CAN. Although, the trends for 50 day results are very similar the uncertainties are slightly overestimated.

Just three methods are used for  $\nu_{\rm max}$. For OCT and CAN there is general agreement between MAD and uncertainty with a slight tendency for the uncertainties to be over estimated. The uncertainties for COR:EACF are overestimated by roughly a factor of two to three.

Additionally, for all methods the variation of MAD with frequency is not strong and supports our earlier assumption for the outlier rejection to use a fixed threshold independent of frequency.


\subsection{Offsets between different methods}
\begin{figure}
\begin{minipage}{\linewidth}
\centering
\includegraphics[width=\linewidth]{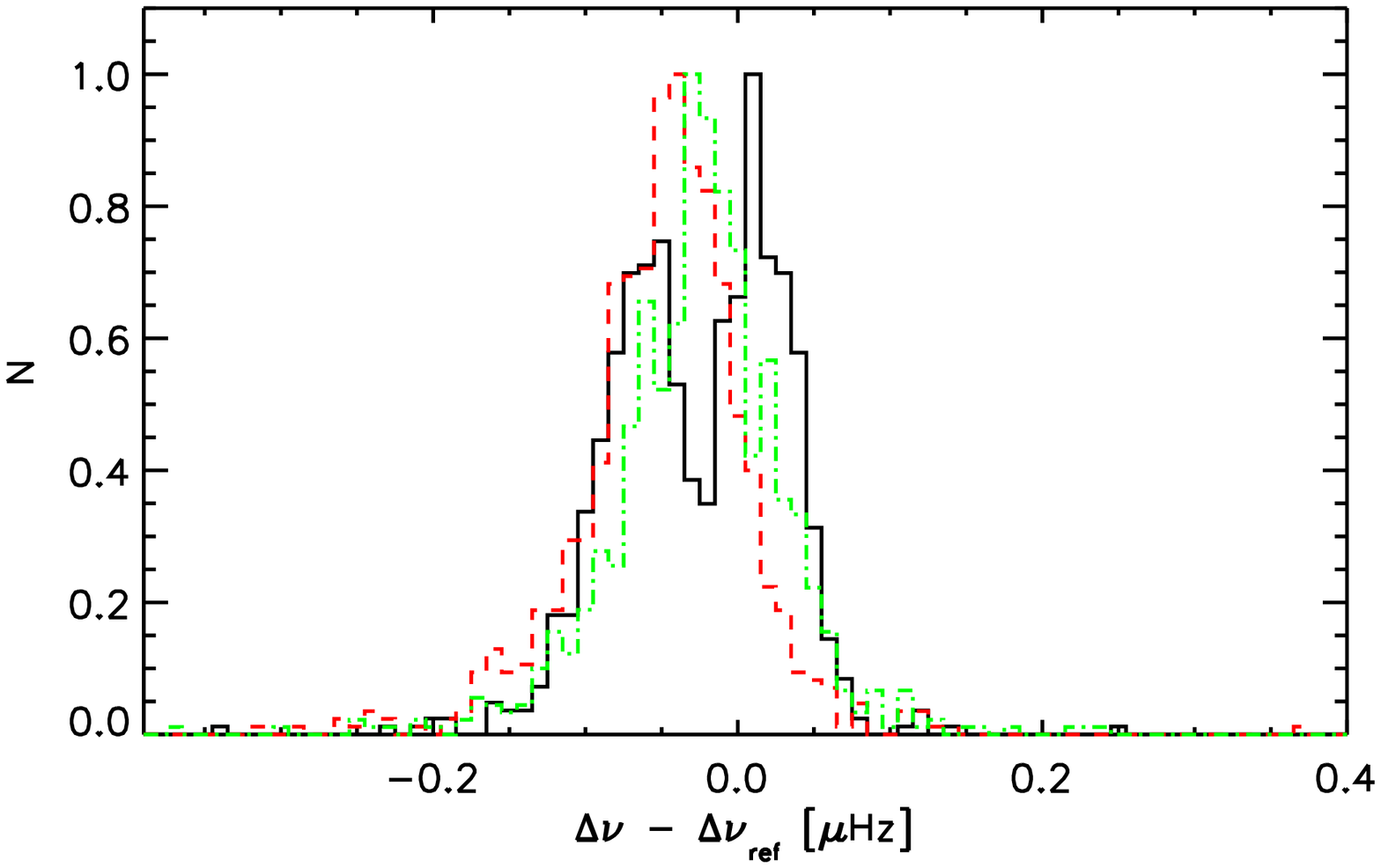}
\end{minipage}
\begin{minipage}{\linewidth}
\centering
\includegraphics[width=\linewidth]{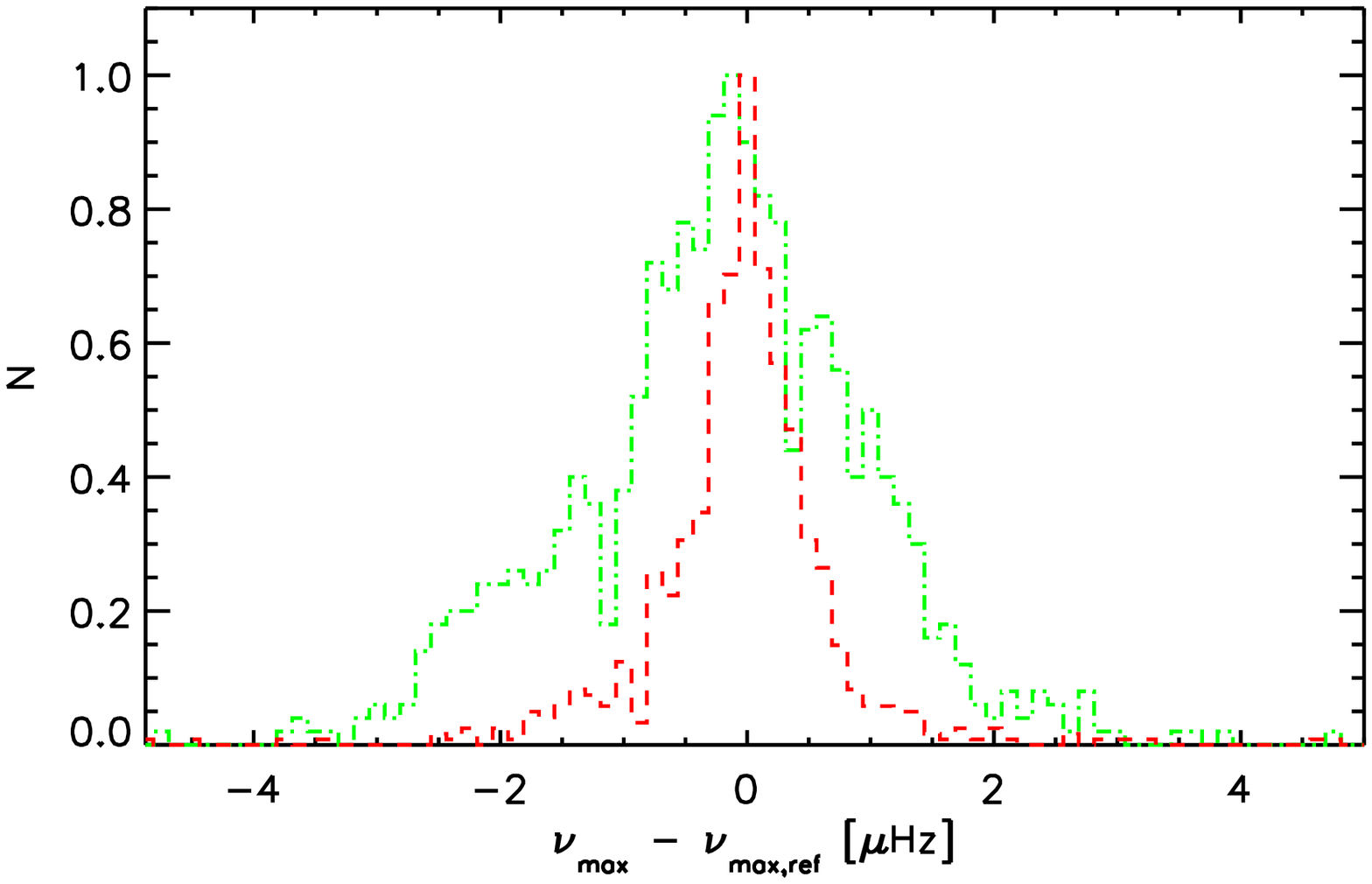}
\end{minipage}
\caption{Normalised distribution of the offset of the individual 600 day results from the reference value, i.e., the CAN 600 day results, for $\meandnu$ (top) and $\nu_{\rm max}$ (bottom). COR:UP, COR:EACF and OCT results are indicated in black solid lines, green dashed-dotted lines and red dashed lines respectively.}
\label{compmeth}
\end{figure}

In the previous subsection we saw that within any one given method, short datasets can give, on average, slightly biased results when compared with longer sets. Here we concentrate on the differences between different methods using the results obtained with 600 days of data. We know that the different methods involve different assumptions and no method is without assumptions as is shown by \citet{kallinger2012}. Two methods can be considered to lie at extreme ends of the choices for how to measure $\meandnu$. At one extreme is CAN which uses individual peak bagging to measure two values of $\Delta\nu$ close to the peak of the oscillation power and returns their average as $\meandnu$.  At the other end of the choice is COR:UP, which imposes a regular pattern on the whole spectral range and returns a $\meandnu$ based on that. It is known that variation of the large separation with frequency is dependent on the evolutionary state of the star \citep{kallinger2012} and this is seen very clearly if the values of $\meandnu$ from CAN and COR:UP are compared (see Fig.~\ref{compmeth}). Indeed the COR:UP show a bimodal distribution with respect to the CAN results, in which the left peak are predominantly RC stars and the rightmost peak are RGB stars. Following the reasoning of \citep{kallinger2012}, this clear difference between $\meandnu$ could even be used to classify whether a star is already in its He-core burning phase. For the other methods the differences follow the same pattern, but are not as clear because, firstly they are in between CAN and COR:UP in terms of their global / local approach and secondly CAN and COR:UP are not particular sensitive to realization noise. For COR:UP this is because of the regularity constraint and for CAN it is due to the fact that the frequency determination of a given peak is relatively insensitive to the realization noise given the long datasets.

For $\nu_{\rm max}$, effects are less pronounced. OCT agrees well with CAN but still with a (small) difference between RGB and RC. This could be due to difference in the acoustic cutoff frequency and / or differences in the smoothing applied to fit the power excess. \citet{mosser2012} investigated this in detail and showed that smoothing can have a non-negligible effect (also already pointed out by \citet{kallinger2010}). Furthermore, they show that clump stars have oscillations with lower amplitudes, but larger $\nu_{\rm max}$, than stars ascending the red-giant branch with similar values for $\meandnu$.

This comparison of results of long datasets obtained with different methods shows that the definition of the obtained parameter is of importance and that the differences in the definition are significantly larger than the observational uncertainties. Hence it is important when quoting a parameter value to provide the detailed definition of that particular parameter. Note that all methods also differ in their sensitivity to realization noise as already seen in paper I.

\subsection{Comparison between the predicted and observed mode $H/B$}
\label{ss:rms}
For each star analyzed, a value for the envelope height and the noise background at  $\nu_{\rm max}$ are returned. We have certain expectations for the values.
We expect, on average, the ratio of these two parameters to be have a value of about 3.7 or 4.0 depending on the evolutionary status of the star \citep{mosser2012}.
From the same work we know that within factors of order unity the values returned by different methods will not be entirely consistent.
In this section we explore how closely the expectations are met.
We also look at how the ratio varies from run to run particularly for the short runs in order to evaluate whether this is a significant factor in the non-detection of the oscillations.
For the longest available dataset of 600 days, the median value of the observed $H/B$ is 4.1 with inter-quartile distance of 1.4 which is roughly consistent with the expectations.

Next we turn to a consideration of the 50-day data. Here we find that on average the returned envelope height and noise background are consistent with the figures for the longer runs. However, this masks a large amount of variability. The apparent height of the envelope is very variable. We do not know if this is genuine variability or a defect in the algorithms.
However, it is clear that even with height-to-background ratios significantly below unity, detection of the modes is possible thanks to the regular pattern of the oscillations.  We do not have the values where the algorithms failed to find evidence for oscillations and so cannot comment on the height-to-background ratio in these cases.

\section{Prediction of $\nu_{\rm max}$ from rms flux}
\begin{figure}
\begin{minipage}{\linewidth}
\centering
\includegraphics[width=\linewidth]{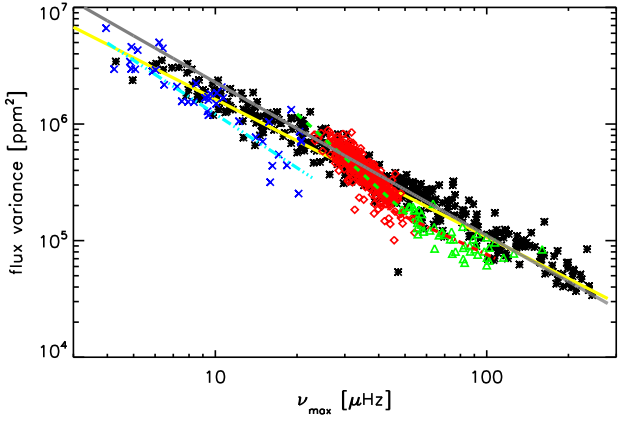}
\end{minipage}
\caption{Variance of the flux as a function of $\nu_{\rm max}$, with RGB, RC, second clump and AGB stars indicated by black asterisks, red diamonds, green triangles and blue crosses, respectively. Fits to the values of the four evolutionary states are shown by the yellow solid line, the green dashed line, the red dashed-dotted line and light-blue dashed-triple dotted line. The prediction from Eq.~\ref{finalpred} is indicated with the gray line.}
\label{nu_maxrms}
\end{figure}

\begin{table}
\begin{minipage}{\linewidth}
\caption{Coefficients of the fit: $V = a\nu_{\rm max}^b$, with $V$ the variance of the flux in ppm$^2$ and $\nu_{\rm max}$ the frequency of maximum oscillation power in $\mu$Hz, for different evolutionary phases.}
\label{tabcof}
\centering
\begin{tabular}{lcc}
\hline\hline
& a & b  \\
\hline
RGB & $2.4\times10^7 \pm 1\times10^6$ & $-1.18 \pm 0.01$\\
RC & $7\times10^8 \pm 2\times10^8$ & $-2.13 \pm 0.07$\\
second clump & $1.1\times10^7 \pm 5\times10^6$ & $-1.1 \pm 0.1$\\
AGB & $4.1\times10^7 \pm 7\times10^6$ & $-1.53 \pm 0.08$\\
\hline
\end{tabular}
\end{minipage}
\end{table}

An automated analysis of the red giant data is made more difficult by the fact that for some of the largest giants the peak in the mode power is at very low frequency (below $\sim$ 5~$\mu$Hz). Unless the datasets are very long, the spectra do not have enough resolution to clearly distinguish the oscillations. The automated algorithms may then fasten on features at other frequencies and thus provide a false positive detection. We therefore have sought an independent parameter to guide the software to the appropriate region. We have found that the mean flux variance in the timeseries data is one such guide.  We first provide an analysis which shows why this should be so and then provide the data to illustrate the dependence that we find.

Parseval's theorem states that the variance of the timeseries is equal to the integrated power in the spectrum. We therefore look at the sources of power in the spectrum.
At very low frequencies, instrumentation effects will become important. To some extent this has been removed from the data considered here by the data preparation algorithms. At all frequencies there is photon shot noise, but the red giants are usually sufficiently bright that it can be neglected.
As a consequence, for red giants the major sources of the signal in the data are the granulation and the oscillations.
The mode power is  modelled as a Gaussian of height $H$ and full width half power  $\delta_{\rm env}$ hence the total power in the modes is
\begin{equation}
P_{\rm mode\_total}=\frac{1}{2}\sqrt{\frac{\pi}{\ln2}}H\delta_{\rm env}.
\end{equation}
Using \citet{mosser2012}, we can express both the height at maximum and the width of the
distribution as a function of the frequency of maximum power:
\begin{equation}
H\delta_{\rm env} = 1.4 \times 10^7  \nu_{\rm max}^{-1.5}.
\end{equation}
The frequency distribution of the power in the granulation is modelled according to
the  Harvey prescription:
\begin{equation}
P_{\rm gran}(\nu) = \frac{4\sigma_{\rm int}^2\tau_{\rm gran}}{1+(2\pi\tau_{\rm gran}\nu)^2},
\end{equation}
where variance in the timeseries of the granulation is $\sigma_{\rm int}^2$ and $\tau_{\rm gran}$ is the timescale of the granulation.
We can use this to estimate the power, $B$, in the granulation signal at $\nu_{\rm max}$.
At $\nu_{\rm max}$ the factor of unity in the denominator can be neglected:
\begin{equation}
B = \frac{\sigma_{\rm int}^2}{\pi^2\nu_{\rm max} \times \tau_{\rm gran}\nu_{\rm max}}.
\end{equation}
From \citet{mathur2011} we have that $\tau_{\rm gran} \approx 0.7 \nu_{\rm max}^{-0.9}$, hence
\begin{equation}
B = \frac{\sigma_{\rm int}^2}{0.7 \pi^2 \nu_{\rm max}^{1.1}}.
\end{equation}
Knowing that $H=2.03 \times 10^7 \nu_{\rm max}^{-2.38}$ we can use the observation that the ratio of height to background is a constant of value 3.7 to 4 depending on the evolutionary state of the star:
\begin{equation}
\frac{H}{B} = \frac{14 \times 10^7 \nu_{\rm max}^{-1.28}}{\sigma_{\rm int}^2}.
\end{equation}
Knowing $H/B$ we can now estimate a value for $\sigma_{\rm int}^2$
Thus the total variance ($V$) in the timeseries is
\begin{equation}
V = \sigma_{\rm int}^2 + P_{\rm mode\_total},
\end{equation}
\begin{equation}
V = \frac{14 \times 10^7 \nu_{\rm max}^{-1.28}}{H/B} + 1.4 \times 10^7\nu_{\rm max}^{-1.5} \rm ~ppm^2.
\end{equation}
A typical value for $H/B$ is about 4, hence
\begin{equation}
V = (3.5\nu_{\rm max}^{-1.28} + 1.4 \nu_{\rm max}^{-1.5}) \times 10^7 \rm ~ppm^2.
\label{finalpred}
\end{equation}
It is clear that although the power law indices of $\nu_{\rm max}$ in the two
components of the noise are not the same, they are relatively close to each other.
The granulation provides just over twice the amount of power as do the modes. 

Observationally we get $2.4 \times 10^7 \nu_{\rm max}^{-1.18}$ ppm$^2$ for RGB stars (see fit in Fig.~\ref{nu_maxrms}). We see that for other evolutionary states the fits have different coefficients (Table~\ref{tabcof}). This indicates that there are differences in either the granulation description and / or the height and width ratio of the oscillation power as a function of evolution phase. This is consistent with what is shown by \citet{mosser2012}, and needs further investigations which is beyond the scope of this paper. 


\section{Summary}

In this work we investigated the impact of the length of the timeseries on the precision and accuracy of the determined global oscillation parameters $\nu_{\rm max}$ and $\meandnu$ of red giants. We used \textit{Kepler} light curves spanning about 600 days and divided them in short runs of 50, 100, 200 and 400 days. All these runs have been analysed using automated methods. The oscillation detection rate has been compared with predictions and the resulting values for the global oscillation parameters have been compared as a function of method, run length, $\meandnu$ of the oscillations.
From this study we find that:
\begin{itemize}
\item For 95\% of the stars consistent global oscillation parameters are obtained from 600 day timeseries with different methods. For the remaining 5\%, there were good reasons for the lack of consistency.  
\item Using the observational methods we find more than 95\% (of the consist results of 600 day data) or more reliable detections of oscillations in timeseries of 400 days or longer.
\item Current predictions of the detectability of oscillations are based on the amplitudes and predict that in the majority of the cases the likelihood to detect oscillations are above 90\% for both the long and short runs. However, most observational algorithms use the regularity in the power spectrum to detect the oscillations and the regularity has reduced sensitivity for shorter runs.
\item The precision of the determined global oscillation parameters increases with increasing timeseriess and the trends suggest that this continues for even longer timeseries than investigated here. From the extrapolation of fits to the median absolute deviations a reduction of more than a factor of 10 for an increase in timespan from 50 to 2000 days (the currently foreseen length of the mission) is foreseen. Thus, there are real advantages to be gained from working with even long timeseries than considered here. We note that the universal pattern is already effective for short datasets.
\item The distributions of the offsets - difference between results of short runs with respect to the result obtained with the same method on the 600-day long timeseries -  divided by the quoted uncertainties show that the quoted uncertainties have a tendency to be overestimated, which is in general more severe for longer datasets. However, this does depend on the method. 
\item We find that 50 day timeseries are not long enough to be certain to pick up more than 90\% of the oscillations with the currently employed methods.
\item When comparing different methods it is clear that the differences due to different definitions are non-negligible. This difference is a function of the evolutionary state of the stars and this could be used to determine the evolutionary state.
\item The different strengths, definitions and sensitivity to realization noise of the different methods indicate that the simultaneous use of more methods is likely to be profitable.
\end{itemize}

Additionally, we propose and justify a new method to estimate the frequency of maximum oscillation power from variance in the timeseries. We show that the dependence of the flux variance on $\nu_{\rm max}$ is also a function of evolutionary phase. The effectiveness of this method does not depend on the data duration nor on the location of the peak of the spectrum -- always assuming that the necessary data detrending is not attenuating the oscillations signal. We recommend that this method be used in conjunction with the methods described here as an additional independent constraint to detect the oscillations.








\acknowledgements
Funding for this Discovery Mission is provided by NASA's Science Mission Directorate. The \textit{Kepler} Team is recognized for helping to make the mission and these data possible. SH acknowledges financial support from the Netherlands Organisation for Scientific Research (NWO). YE and WJC acknowledge support from the Science and Technology Facilities Council (STFC).
\bibliographystyle{aa}
\bibliography{comparison2}

\end{document}